\documentclass[aps,prl,twocolumn,showpacs,superscriptaddress,groupedaddress]{revtex4}  
\usepackage{graphicx}  
\usepackage{dcolumn}   
\usepackage{bm}        
\usepackage{amssymb}   
\usepackage{mathrsfs}

\def\avg#1{\big <#1 \big >}
\def\diamondcomma{\ \raise.3ex\hbox{$\diamond$}\kern-0.4em\lower.7ex\hbox{$,$}\ }
\def\lesssim{\ \raise.3ex\hbox{$<$}\kern-0.8em\lower.7ex\hbox{$\sim$}\ }
\def\gesim{\ \raise.3ex\hbox{$>$}\kern-0.8em\lower.7ex\hbox{$\sim$}\ }

\begin{document}
\widetext
\title{Non-Hermitian phase transition from a polariton Bose-Einstein condensate to a photon laser}
\author{Ryo Hanai}
\email{hanai@acty.phys.sci.osaka-u.ac.jp}
\affiliation{James Franck Institute and Department of Physics, University of Chicago, Illinois, 60637, USA} 
\affiliation{Department of Physics, Osaka University, Toyonaka 560-0043, Japan} 
\author{Alexander Edelman}
\affiliation{James Franck Institute and Department of Physics, University of Chicago, Illinois, 60637, USA}
\affiliation{Materials Science Division, Argonne National Laboratory, Argonne, Illinois 60439, USA} 
\author{Yoji Ohashi}
\affiliation{Department of Physics, Keio University, Yokohama 223-8522, Japan} 
\author{Peter B. Littlewood}
\affiliation{James Franck Institute and Department of Physics, University of Chicago, Illinois, 60637, USA} 
\affiliation{Materials Science Division, Argonne National Laboratory, Argonne, Illinois 60439, USA} 
\date{\today}
\begin{abstract}
We propose a novel mechanism for a nonequilibrium phase transition in a $U(1)$-broken phase of an electron-hole-photon system, from a Bose-Einstein condensate of polaritons to a photon laser, induced by the non-Hermitian nature of the condensate.
We show that a (uniform) steady state of the condensate can always be classified into two types, namely, arising either from lower or upper-branch polaritons. We prove (for a general model) and demonstrate (for a particular model of polaritons) that an exceptional point where the two types coalesce marks the endpoint of a first-order-like phase boundary between the two types, similar to a critical point in a liquid-gas phase transition.
Since the phase transition found in this paper is not in general triggered by population inversion, our result implies that the second threshold observed in experiments is not necessarily a strong-to-weak-coupling transition, contrary to the widely-believed understanding.
Although our calculation mainly aims to clarify polariton physics, 
our discussion is applicable to general driven-dissipative condensates composed of two complex fields. 
\end{abstract}
\maketitle

The phenomenon of macroscopic condensation has been one of the principal topics in modern condensed matter physics and optics \cite{Proukakis}.
The central example is, of course, Bose-Einstein condensation (BEC), which has been observed in various systems, ranging from atomic gases \cite{Anderson1995,Davis1995}, liquid ${}^4$He \cite{Huang}, exciton-polaritons \cite{Kasprzak2006,Deng2010,Carusotto2013,Byrnes2014}, magnons \cite{Ruegg2003,Demokritov2006,Chumak2009}, photons \cite{Klaers2010}, to plasmonic-lattice-polaritons \cite{Hakala2018}.
In these systems, thermalization plays a crucial role in achieving macroscopic occupation of the lowest energy level.
A photon laser \cite{Scully,Haug}, in contrast, is a nonequilibrium condensate, where the 
population inversion in an optical gain medium induces macroscopic coherence.

The semiconductor microcavity system \cite{Kasprzak2006,Deng2010,Carusotto2013,Byrnes2014} 
provides a unique opportunity to study similarities and differences of these two classes of condensation phenomena \cite{Imamoglu1996}, 
since it can exhibit both \cite{Deng2003}, by tuning the pump power.
At low pump power, where the strong light-matter coupling enables hybrid light-matter quasiparticles called polaritons to form,
their thermalization is efficient due to relaxation processes such as stimulated scattering. 
This makes it possible, once the pump power exceeds a certain threshold, for the system to exhibit macroscopic coherence among polaritons to turn into a polariton-BEC \cite{Kasprzak2006}.
At even higher power, in contrast, the system operates in the weak light-matter coupling regime as a vertical-cavity surface-emitting laser (VCSEL), a type of a photon laser, with electrons and holes acting as a gain medium.
Interestingly, a number of experiments \cite{Bajoni2008,Balili2009,Nelsen2009,Tempel2012a,Tempel2012b,Tsotsis2012,Horikiri2013,Fischer2014,Kim2016,Brodbeck2016,Dietrich2016,Schneider2013} have observed a second threshold between the former and latter regimes, which has been traditionally interpreted as 
a strong-to-weak coupling phase transition.

This two-threshold-behavior presents a theoretical challenge, however.
The normal-to-lasing transition is associated with breaking a $U(1)$ symmetry, but the polariton-BEC 
is already in a $U(1)$-broken phase. Thus, there seems to be no good reason to expect a second phase transition.
Indeed, to our knowledge, all theories to date predict a crossover \cite{Szymanska2006,Szymanska2007,Yamaguchi2012,Yamaguchi2013,Yamaguchi2015}. 

\begin{figure}
\begin{center}
\includegraphics[width=1\linewidth,keepaspectratio]{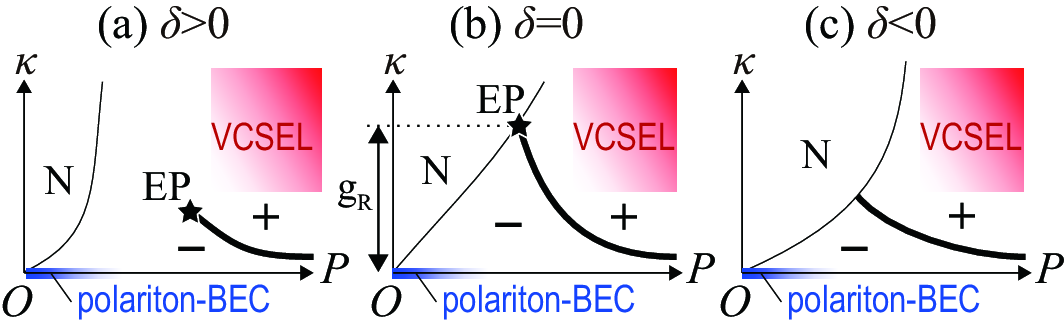}
\end{center}
\caption{(Color online) 
Proposed phase diagram of a driven-dissipative electron-hole-photon gas, in terms the photon decay rate $\kappa$ and the pump power $P$. (a) Blue detuning. (b) On resonance. (c) Red detuning. 
``$-(+)$'' represents the ``$-(+)$''-solution phase, ``N'' represents the normal phase, ``EP'' is the exceptional point, and
$g_{\rm R}$ is the Rabi splitting.
The thick (thin) solid line represents the phase boundary in the condensed phase (between the normal and the condensed phase).}
\label{fig_phasediagram}
\end{figure}

In this Letter, we propose a novel mechanism for a phase transition in the $U(1)$-broken phase, triggered by the non-Hermitian nature of the out-of-equilibrium condensate.
Starting from the equation of motion of a microscopic model, 
we show that the steady states of a two-component condensate of electron-hole pairs and photons can formally be classified into two types of solutions, corresponding to condensation into different branches of the polariton spectrum. 
We find that an exceptional point (EP), where the two solutions coalesce \cite{Kato1960,Bender1998,Heiss1999,Dembowski2004,Heiss2012,Dembowski2005,Lee2009,Gao2015}, may appear due to the non-Hermiticity of the equation of motion. 
We prove and demonstrate that this is the endpoint of a first-order-like phase transition line between the two solutions, analogous to a critical point in a liquid-gas phase diagram.
Based on these results, we propose a phase diagram of an electron-hole-photon system depicted in Fig. \ref{fig_phasediagram}.
Our theory points out the possibility of both the crossover and phase transition from polariton-BEC to VCSEL depending on the experimental settings such as detuning and the pump power, 
and provides  a possible new interpretation to the second threshold as a signal of a lower to upper branch transition.
These physics, although derived mainly with microcavity polaritons in mind, should be applicable to other driven-dissipative many-body systems with coupled order parameters, e.g.
atoms in a double-well potential \cite{Graefe2012,Cartarius2012,Dast2013}, a supersolid realized in two-crossed cavity \cite{Leonard2017}, 
or a plasmonic-lattice-polariton BEC \cite{Hakala2018}.

\begin{figure}
\begin{center}
\includegraphics[width=0.8\linewidth,keepaspectratio]{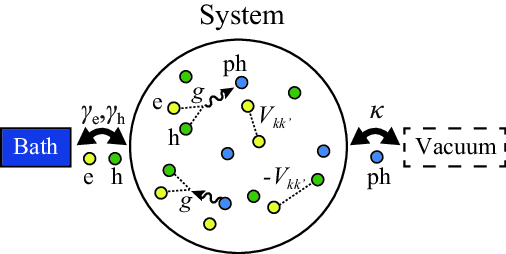}
\end{center}
\caption{(Color online) Model driven-dissipative electron-hole-photon gas. The system is attached to an electron-hole bath and a photon vacuum.
Electrons (holes) are incoherently supplied to the system with the rate $\gamma_{\rm e(h)}$. 
In the system, the injected electrons (``e'') and holes (``h'') repulsively (e-e, h-h) and attractively (e-h) interact with the Coulomb potential $V_{\bm k-\bm k'}=e^2/(2\epsilon|\bm k-\bm k'|)$. The electrons and holes pair-annihilate (create) to create (annihilate) cavity-photons (``ph'') via the dipole coupling $g$.
The created photons in the cavity leak out to the vacuum with the decay rate $\kappa$.
}
\label{fig_model}
\end{figure}

We use a microscopic model schematically shown in Fig. \ref{fig_model} \cite{Yamaguchi2012,Yamaguchi2013,Yamaguchi2015,Hanai2018}, which has been shown to capture both the essential physics of the BEC state and the VCSEL \cite{NOTE_Yamaguchi}, as well as to give a  semiquantitative agreement \cite{Hanai2018} with photoluminescence experiments \cite{Assmann2011,Tempel2012a,Nakayama2016,Nakayama2017}. The system is composed of electrons, holes, and cavity photons, which are coupled to an electron-hole bath and a photon vacuum.
Electrons (holes) are incoherently pumped to the system from the bath at a rate $\gamma_{\rm e(h)}$. 
The injected electrons and holes Coulomb-interact with each other and create (annihilate) photons by pair-annihilation (creation).
The photons leak out to the vacuum with the decay rate $\kappa$, driving the system into a non-equilibrium steady state.
The explicit expression for the Hamiltonian $H$ is given in the Supplemental Material (SM) \cite{Supp}.

We apply the Keldysh Green's function method \cite{Rammer} to the model.
As shown in SM \cite{Supp}, the dynamics of the electron-hole dipole polarization $p_{\bm k}(\bm r,t)$ and the electron (hole) density $n_{\bm k,\sigma={\rm e(h)}}(\bm r,t)$ obeys the generalized Boltzmann equation \cite{Baym},
\begin{eqnarray}
&&i\hbar\partial_t p_{\bm k}(\bm r,t)
=
\Big[
\varepsilon_{\bm k,{\rm e}}
+\varepsilon_{\bm k,{\rm h}}
-\frac{\hbar^2\nabla^2}{4m_{\rm eh}}
-2i\gamma
\Big]
p_{\bm k}(\bm r,t)
\nonumber\\
&&\ \ \ \ \ \ \ \ \ \ \ \ \ \ \ 
-\sum_{\bm k'}L_{\bm k,\bm k'}(\bm r,t)\Delta_{\bm k'}(\bm r,t),
\label{pk}
\\
&&\partial_t n_{\bm k,\sigma}(\bm r,t)
+\bm v_{\bm k,\sigma}\cdot \nabla n_{\bm k,\sigma}(\bm r,t)
\nonumber\\
&&\ \ \ \ \ \ \ \ \ \ \ \ \ \ \ 
=
-\frac{2\gamma_\sigma}{\hbar} n_{\bm k,\sigma}(\bm r,t)
+I_{\bm k,\sigma}(\bm r,t).
\label{nk}
\end{eqnarray}
Here, $\varepsilon_{\bm k,{\rm e(h)}}=\hbar^2\bm k^2/(2m_{\rm e(h)})+E_{\rm g}/2$
is the dispersion of the electron (hole) in the conduction (valence) band, where $m_{\rm e(h)}$ is the effective mass of electrons (holes). $E_{\rm g}$ is the energy gap of the semiconductor material.  
$m_{\rm eh}=2m_{\rm e}m_{\rm h}/(m_{\rm e}+m_{\rm h})$ is twice the reduced mass of an electron and a hole, and $\bm v_{\bm k,{\rm e(h)}}=\hbar\bm k/m_{\rm e(h)}$.
We have introduced the order parameter $\Delta_{\bm k}(\bm r,t)=\sum_{\bm k'}V_{\bm k-\bm k'}p_{\bm k'}(\bm r,t)-g\lambda_{\rm cav}(\bm r,t)$ describing the condensed phase, where $\lambda_{\rm cav}(\bm r,t)=\avg{a(\bm r,t)}$ is the coherent cavity-photon amplitude (where $a(\bm r,t)$ is the annihilation operator of a cavity-photon), $V_{\bm k}=e^2/(2\epsilon|\bm k|)$ is the two-dimensional Coulomb interaction ($\epsilon$ is the dielectric constant), and $g$ is a dipole coupling between carriers (electrons and holes) and photons.
The coupling of the system to the bath causes the dephasing/decay of $p_{\bm k}(\bm r,t)$ ($n_{\bm k,\sigma}(\bm r,t)$) with the rate $2\gamma$ ($2\gamma_\sigma$), 
where $\gamma=(\gamma_{\rm e}+\gamma_{\rm h})/2$.
$L_{\bm k,\bm k'}(\bm r,t)$ and $I_{\bm k,\sigma}(\bm r,t)$  in Eqs. (\ref{pk}) and (\ref{nk}), determined microscopically from the self-energy $\hat\Sigma$ and the Green's function $\hat G$ in the Nambu-Keldysh formalism (see SM \cite{Supp} for their explicit form), 
describe many-body interaction effects such as exciton formation, collision, phase-filling, etc., as well as the electron-hole pumping and its thermalization.

The electron-hole dynamics is coupled to the dynamics of the coherent cavity-photon amplitude, given by the Heisenberg equation \cite{Supp},
\begin{eqnarray}
i\hbar\partial_t\lambda_{\rm cav}(\bm r,t)
&=&
\avg{[a(\bm r,t),H]}
=
\Big[
\hbar\omega_{\rm cav}
-\frac{\hbar^2\nabla^2}{2m_{\rm cav}}
-i\kappa
\Big]
\nonumber\\
&\times& \lambda_{\rm cav}(\bm r,t) 
+g \sum_{\bm k}p_{\bm k}(\bm r,t),
\label{Langevin} 
\end{eqnarray}
where $\hbar\omega_{\rm cav}$ is the cavity-photon energy, and $m_{\rm cav}$ is a cavity-photon mass. 
In analogy to $\lambda_{\rm cav}(\bm r,t)$, we define for later use a complex electron-hole pair amplitude $\lambda_{\rm eh}(\bm r,t)$ by $p_{\bm k}(\bm r,t) = \lambda_{\rm eh}(\bm r,t)\phi_{\bm k}(\bm r,t)$, $\sum_{\bm k}|\phi_{\bm k}(\bm r,t)|^2=1$, ${\rm Arg}[\sum_{\bm k}\phi_{\bm k}(\bm r,t)]=0$ \cite{Comte1982}.

Our main assumption in what follows is that the system supports spatially uniform, steady-state solutions given by the ansatz \cite{Szymanska2006,Szymanska2007,Yamaguchi2012,Yamaguchi2013,Yamaguchi2015,Hanai2018,Hanai2017,Hanai2016} 
$\lambda_{\rm cav(eh)}(t)=\lambda_{\rm cav(eh)}^0 e^{-iEt/\hbar}$, 
where $E$ is the (real) condensate emission energy.
Although, in real systems, there is always a chance that such uniform steady state destabilizes, e.g. due to the dynamical instability that leads to pattern formation \cite{Daskalakis2015,Bobrovska2018,Baboux2018} 
or the occurence of many-body localization \cite{Sturges2019},
we ignore such possibilities in this Letter. 
In this formulation, $\lambda^0_{\rm cav(eh)}$ corresponds to the photonic (excitonic) component of the macroscopic many-body wave function.

With this ansatz, Eqs. (\ref{pk}) and (\ref{Langevin}) satisfies a non-Hermitian eigenvalue equation,
\begin{eqnarray}
\hat A 
\left(
\begin{array}{c}
\lambda_{\rm cav}^0 \\
\lambda_{\rm eh}^0
\end{array}
\right)
=
\left(
\begin{array}{cc}
h_{\rm cav} & g_0  \\
\tilde g_0^* & h_{\rm eh}
\end{array}
\right)
\left(
\begin{array}{c}
\lambda_{\rm cav}^0 \\
\lambda_{\rm eh}^0
\end{array}
\right)
=E\left(
\begin{array}{c}
\lambda_{\rm cav}^0 \\
\lambda_{\rm eh}^0
\end{array}
\right),
\label{lambdaalpha}
\end{eqnarray}
where $h_{\rm cav}=\hbar\omega_{\rm cav}-i\kappa$, $g_0 =g\sum_{\bm k}\phi_{\bm k}$, 
$\tilde g_0^*=g\sum_{\bm k,\bm k'}\phi^*_{\bm k}L_{\bm k,\bm k'}$, and $h_{\rm eh} =\sum_{\bm k}[(\varepsilon_{\bm k,{\rm e}}+\varepsilon_{\bm k,{\rm h}}-2i\gamma)|\phi_{\bm k}|^2-\sum_{\bm p,\bm k'}V_{\bm k-\bm p}\phi_{\bm k}^*\phi_{\bm p}L_{\bm k,\bm k'}]$.
We emphasize that Eq. (\ref{lambdaalpha}) is a steady state condition that determines the macroscopic variables $\lambda^0_{\rm cav(eh)}$ and is analogous to a gap equation, not to be confused \cite{NOTE_polariton} with the equations for determining the polariton spectra in the normal state \cite{Deng2010}.
For instance, the trivial solution $\lambda^0_{\rm cav} = \lambda^0_{\rm eh} = 0$ describes the normal state.

Eqs. (\ref{pk})-(\ref{Langevin}) must be solved self-consistently for a given set of microscopic parameters to determine the quantities that enter Eq. (\ref{lambdaalpha}) \cite{NOTE_solve}. However, we can draw a number of strong conclusions by analyzing the structure of the latter alone. 
The matrix $\hat A$ can be diagonalized with eigenvectors
$\bm u_- = 
   (\frac{-\varphi +\Omega}{2}, -\tilde g_0^*)^{\mathsf T},\
\bm u_+ = (g_0 , \frac{-\varphi +\Omega}{2})^{\mathsf T},$
and corresponding eigenvalues $E_\pm=[h_{\rm cav}+h_{\rm eh} \pm \Omega]/2$.
Here, $\Omega=\sqrt{\varphi^2 + 4\tilde g_0^*g_0}$, $\varphi=h_{\rm cav} - h_{\rm eh}$, and we take ${\rm Re}\Omega\ge 0$ (i.e. ${\rm Re}E_+\ge {\rm Re}E_-$) without loss of generality.
In the diagonal basis, Eq. (\ref{lambdaalpha}) reads 
$(E_- - E)\lambda_-^0=(E_+ - E)\lambda_+^0=0$,
where 
$(\lambda_-^0,\lambda_+^0)^{\mathsf T}
=\hat U(\lambda_{\rm cav}^0,\lambda_{\rm eh}^0)^{\mathsf T}$ 
with 
$\hat U^{-1}=(\bm u_-,\bm u_+)$.
From this relation, we see that 
$\lambda_-^0$ and $\lambda_+^0$ cannot be non-zero simultaneously as long as $E_-\ne E_+$, allowing us to classify the non-trivial solutions into two types:
$(\lambda_-^0 \ne 0,\lambda_+^0=0,E=E_-)$ and $(\lambda_+^0\ne0,\lambda_-^0=0 ,E=E_+)$, which we label ``$-$'' and ``$+$'', respectively.
This property is essentially different from similar time-dependent coupled-damped oscillators equations, $i\partial_t (\psi_1,\psi_2)^{\mathsf T}=\hat H_{\rm cdo}(\psi_1,\psi_2)^{\mathsf T}$ (where $\psi_1$ and $\psi_2$ are complex numbers and $\hat H_{\rm cdo}$ is a non-Hermitian $2\times 2$ matrix), 
which are often discussed in the field of non-Hermitian quantum mechanics
\cite{Bender1998,Heiss1999,Dembowski2004,Heiss2012,Dembowski2005,Lee2009,Gao2015}, where the transient dynamics generally allows for a superposition of eigenmodes.

\begin{figure}
\begin{center}
\includegraphics[width=0.92\linewidth,keepaspectratio]{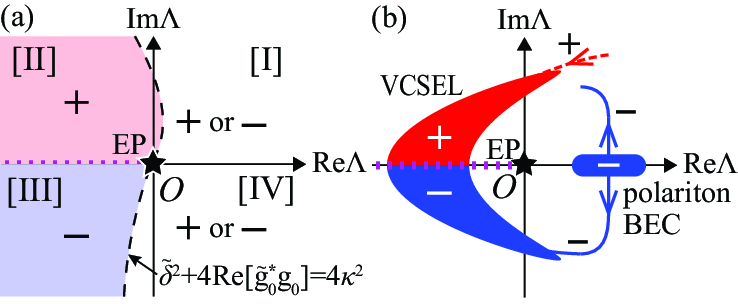}
\end{center}
\caption{(Color online) 
(a) 
Definition of regions I-IV.
In region II (III) in the weak-coupling regime, only the ``$+(-)$''-solution is allowed. 
On the dotted line, the solution type switches without being accompanied by discontinuity. 
(b) Schematic description of how a polariton-BEC evolves to a VCSEL, in terms of $\Lambda$.  
The system exhibits a phase transition (crossover) from a polariton-BEC to a VCSEL when $\Lambda$ changes counter-clockwise (clockwise) around EP.
}
\label{fig_Lambda}
\end{figure}

Now we show our main result of this Letter:
A first-order-like phase transition between the two solutions can occur and the exceptional point (EP) $\Omega = 0$, where $\bm u_\pm$ coalesce such that $\hat A$ only has a single eigenvector, marks the endpoint of the phase boundary.
The proof is presented in SM \cite{Supp} and we sketch the argument here.
Introducing the complex splitting between $E_-$ and $E_+$,
\begin{eqnarray}
\Lambda\equiv\Omega^2=\varphi^2+4\tilde g_0^* g_0,
\label{Lambda}
\end{eqnarray}
we divide the complex $\Lambda$-plane into the regions I-IV, 
according to the strong-coupling condition \cite{Savona1995} $\tilde\delta^2+4{\rm Re}[\tilde g_0^*g_0]\ge 4\kappa^2$ (where $\tilde\delta={\rm Re}\varphi$) and the sign of ${\rm Im}\Lambda$, 
as shown in Fig. \ref{fig_Lambda}(a) \cite{NOTE_weakstrong}.
Due to the restriction of real $E$, only one solution type can exist in the weak-coupling regime (regions II and III), which switches label with no physical discontinuity between regions II and III. 
On the other hand, both (distinct) solution types may coexist in the strong-coupling regions I and IV.
Thus, starting from the ``$-$''-solution in region III, while no discontinuity would be seen when entering region II directly, changing parameters in a route that encircles the EP (III$\rightarrow$IV$\rightarrow$I$\rightarrow$II) requires a phase transition in order to end up in the required ``$+$''-solution in region II, proving the result \cite{NOTE_theorem}.

To make contact between the above general arguments and real physical systems, we explicitly solve for the polariton-BEC and VCSEL.
In the dilute equilibrium limit ($\kappa=0,\gamma\rightarrow 0^+,n_{\bm k,\sigma}\ll 1$) where the polariton-BEC is realized, Eq. (\ref{lambdaalpha}) reduces to \cite{Supp}
\begin{eqnarray}
\hat A_{\rm BEC}=
   \left (
      \begin{array}{cc}
        \hbar\omega_{\rm cav} & g_{\rm R} \\
        g_{\rm R}^* & \hbar\omega_{\rm X} 
      \end{array}
    \right ),
    \label{ABEC}
\end{eqnarray}
in the Hartree-Fock-Bogoliubov approximation (HFBA) \cite{Yamaguchi2012,Yamaguchi2013,Yamaguchi2015,Hanai2018}, which is justified in this limit
\cite{NOTE_BEC}.
Here, $\hbar\omega_{\rm X}=E_{\rm g}-E_{\rm X}^{\rm bind}$ is the exciton energy ($E_{\rm X}^{\rm bind}$ is the exciton binding energy)
and $g_{\rm R}=g\phi_{\rm X}(\bm r=0)$ is the Rabi splitting, where $\phi_{\rm X}(\bm r)$ is an exciton wave function obeying the Schr{\"o}dinger equation $\int d\bm r' [-\delta(\bm r-\bm r')\hbar^2\nabla'^2/m_{\rm eh}-V(\bm r-\bm r')]\phi_{\rm X}(\bm r')=-E_{\rm X}^{\rm bind}\phi_{\rm X}(\bm r)$ \cite{NOTE_BEC}.
The eigenvalues, given by 
$E_\pm^{\rm BEC}=[\hbar\omega_{\rm cav}+\hbar\omega_{\rm X}\pm\sqrt{\delta^2+4|g_{\rm R}|^2}]/2$, are just the lower and upper polariton energies \cite{Deng2010}  
(where $\delta = \hbar\omega_{\rm cav}-\hbar\omega_{\rm X}$ is the conventional detuning parameter).
Comparison of the free energies of the two solutions tells us that the ``$-$''-solution always emerges.

When the photon decay rate $\kappa$ is turned on, a phase transition can occur.
In the so-called polariton laser regime, where the gas is dilute enough to maintain the polariton picture, the equation of motion is governed by the driven-dissipative Gross-Pitaevskii (ddGP) equation \cite{Wouters2007} generalized to the two-component case, given by \cite{Supp},
\begin{eqnarray}
\hat A_{\rm GP}=
\left(
\begin{array}{cc}
\hbar\omega_{\rm cav}-i\kappa & g_{\rm R}  \\
g_{\rm R}^* & \hbar\omega_{\rm X}+U_{\rm X}|\lambda_{\rm eh}^0|^2 + i R_{\rm X}
\end{array}
\right),
\label{GP}
\end{eqnarray} 
where $U_{\rm X}$ is an exciton-exciton interaction strength and $R_{\rm X}>0$ describes the net gain of exciton coherence that feeds the condensate \cite{NOTE_GP}, arising microscopically from processes such as stimulated scattering.
This gives 
$E_\pm^{\rm GP}=[\hbar\omega_{\rm cav}+\hbar\omega_{\rm X}+U_{\rm X}|\lambda_{\rm eh}^0|^2-i(\kappa-R_{\rm X})
\pm\Omega_{\rm GP}]/2$ with
$\Omega_{\rm GP}=\sqrt{\tilde\delta^2+4|g_{\rm R}|^2-(\kappa+R_{\rm X})^2
-2i\tilde\delta(\kappa+R_{\rm X})},$
where 
$\tilde\delta=\hbar\omega_{\rm cav}-(\hbar\omega_{\rm X}+U_{\rm X}|\lambda_{\rm eh}^0|^2)$
is an effective detuning that takes into account the Hartree shift of the exciton component.
One finds an EP ($\Omega_{\rm GP}=0$) at $\tilde\delta=0$ and $g_{\rm R}=R_{\rm X}=\kappa$, 
giving rise to a phase transition in its vicinity.

\begin{figure}
\begin{center}
\includegraphics[width=0.67\linewidth,keepaspectratio]{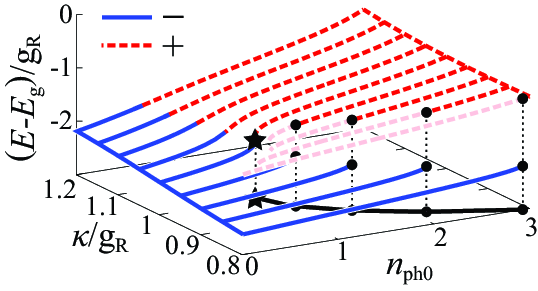}
\end{center}
\caption{(Color online) 
Calculated emission energy $E$ in the case $\hat A=\hat A_{\rm GP}$ as a function of the photon decay rate $\kappa/g_{\rm R}$ and the (coherent) photon number $n_{\rm ph}^0=|\lambda_{\rm cav}^0|^2$.
The solid line projected onto the $n_{\rm ph}^0$-$\kappa/g_{\rm R}$ plane is a phase boundary.
The star represents the EP. 
We set $\delta/g_{\rm R}=0.1,\hbar\omega_{\rm X}/g_{\rm R}=-2,U_{\rm X}/g_{\rm R}=0.1$.}
\label{fig_emission}
\end{figure}

We demonstrate this by explicitly solving Eq. (\ref{lambdaalpha}) when $\hat A=\hat A_{\rm GP}$.
Figure \ref{fig_emission} shows the calculated emission energy $E$ as a function of the decay rate $\kappa$ and the coherent photon number $n_{\rm ph}^0=|\lambda_{\rm cav}^0|^2$ (which roughly corresponds to the pump power), in the blue detuning case $\delta/g_{\rm R}=0.1$.
At $\kappa<g_{\rm R}$, we find that the ``$-$''-solution disappears at a critical value of the pump power, resulting in a phase transition signaled by the discontinuity in $E$.
In constructing the phase diagram, we have assumed that we always realize the lowest-energy solution. Relaxing this assumption would shift the position of the phase boundary in detail but not its endpoint.
As expected, the phase boundary ends at the EP (where $\kappa=g_{\rm R}$). 
When $\kappa>g_{\rm R}$, the ``$-$''-solution crosses over to the ``$+$''-solution.
The fact that a phase transition arises within the ddGP 
(where the polariton picture still holds) suggests that the second threshold observed in experiments does not necessarily imply a strong-to-weak-coupling transition to a photon laser. 
More discussion on this aspect can be found in the SM \cite{Supp}.

At high pump power where the system operates as a VCSEL, 
it has been shown within the HFBA \cite{Yamaguchi2012,Yamaguchi2013,Yamaguchi2015} that Eqs. (\ref{pk})-(\ref{Langevin}) reduce to the semiconductor Maxwell-Bloch equations \cite{Haug}, with
$L_{\bm k,\bm k'}=\delta_{\bm k,\bm k'}N_{\bm k}
=\delta_{\bm k,\bm k'}(1-n_{\bm k,{\rm e}}-n_{\bm k,{\rm h}})$ and
\begin{eqnarray}
\hat A_{\rm VL}=
\left(
\begin{array}{cc}
\hbar\omega_{\rm cav}-i\kappa & g_0  \\
\tilde g_0^{\rm VL*}
& \hbar\omega_{\rm eh}^{\rm VL}-2i\gamma
\end{array}
\right),
\end{eqnarray}
where $\hbar\omega_{\rm eh}^{\rm VL}
=\sum_{\bm k}[(\varepsilon_{\bm k,{\rm e}}+\varepsilon_{\bm k,{\rm h}})|\phi_{\bm k}|^2 
- \sum_{\bm p}V_{\bm k-\bm p}\phi_{\bm k}^*\phi_{\bm p}N_{\bm k}]$ and $\tilde g_0^{\rm VL*}=g\sum_{\bm k}\phi_{\bm k}^*N_{\bm k}$.
A crucial difference compared to the polariton laser case, Eq. (\ref{GP}), is the condensate feeding mechanism.   
The electron-hole gain $R_{\rm X}(>0)$ present in the polariton laser is absent in the VCSEL, 
since the thermalization process does not work efficiently.
Instead, the condensate is fed by stimulated emission arising from the population inversion $N_{\bm k}<0$.
As a result, it is straightforward to show \cite{Supp} that 
${\rm Re}\Lambda_{\rm VL}<0$ holds when ${\rm Im}\Lambda_{\rm VL}= 0$ in the weak-coupling regime \cite{NOTE_VCSEL}, allowing both the solution types to appear and smoothly switch labels with one another.

Figure \ref{fig_Lambda}(b) summarizes the above discussion in terms of 
the complex splitting $\Lambda$.
Here, the polariton-BEC regime lies on the real axis $\Lambda_{\rm BEC}=\delta^2+|g_{\rm R}|^2>0$.
Thus, starting from the polariton-BEC with ``$-$''-solution, by changing parameters such that $\Lambda$ evolves clockwise or counter-clockwise around the EP, the system exhibits a crossover or phase transition, respectively, into a VCSEL.

We connect our discussion in $\Lambda$-space to the physical phase diagram in Fig. \ref{fig_phasediagram}.
Starting from the polariton-BEC ($\kappa=0$), as the decay rate $\kappa$ is turned on such that the system turns into a polariton laser (Eq. (\ref{GP})), 
one sees from the expression of $\Lambda_{\rm GP}=\Omega_{\rm GP}^2$ that
${\rm Im}\Lambda$ increases (decreases) from zero in the case of an effective red (blue) detuning $\tilde\delta<0$ ($>0$), 
where $\Lambda$ evolves counter-clockwise (clockwise).
Since the increasing pump power $P$ usually shifts the effective detuning to red 
(note that $\tilde\delta=\delta-U_{\rm X}|\lambda_{\rm eh}^0|^2$), we predict that there always exists a phase boundary between the polariton-BEC and VCSEL in red detuning, $\delta<0$ [panel (c)]. 
On the other hand, in blue detuning, $\delta>0$, $\tilde\delta$ may switch its sign to negative when $P$ increases. 
Whether this sign change occurs at a positive or negative ${\rm Re}\Lambda$ determines whether the evolution of $\Lambda$ may reverse to counter-clockwise.
Thus, we conjecture that, in the blue detuning case, there exists a phase boundary with an endpoint, as shown in panel (a).
On resonance, $\delta=0$, since we know from Eq. (\ref{GP}) that the EP is at $\kappa=g_{\rm R}$ in the dilute limit $|\lambda_{\rm eh}^0|\rightarrow 0$ 
($\tilde\delta=\delta=0$), the EP lies on the boundary between the normal and the condensed phase [panel (b)].

Physically, when the effective detuning becomes more red, the lower branch becomes more photonic \cite{Deng2010}, hindering condensation to the lower branch as photonic losses increase and gain from the excitonic component becomes small. 
Meanwhile, the upper branch becomes more excitonic, which makes the system favor the latter and eventually driving the phase transition.
In contrast, as long as the system stays in effective blue detuning, it remains in the ``$-$''-solution, exhibiting a crossover.

We close our Letter by commenting on the connection to experiments. 
Most reported experiments exhibiting the two-threshold-behavior are done on resonance or in red detuning with a small decay rate $\kappa<g_{\rm R}$ \cite{Bajoni2008,Balili2009,Nelsen2009,Tempel2012a,Tempel2012b,Tsotsis2012,Horikiri2013,Schneider2013,Fischer2014,Kim2016,Brodbeck2016}, while a single-threshold-behavior to a photon laser has been observed at a large blue detuning \cite{Deng2003}. These results are consistent with our proposal (more detailed discussion is provided in SM \cite{Supp}) which makes us hopeful that an experimental encirclement of the EP is within reach.

We thank S. Diehl, D. Myers, S. Mukherjee, M. Yamaguchi, K. Kamide, T. Ogawa, and K. Asano for discussions.  This work was supported by KiPAS project in Keio University. RH was supported by a Grand-in-Aid for JSPS fellows (Grant No. 15J02513). YO was supported by Grant-in-Aid for Scientific Research from MEXT and JSPS in Japan (No. JP18K11345, No. JP18H05406, No. JP16K05503). Work at Argonne National Laboratory is supported by the U. S. Department of Energy, Office of Science, BES-MSE under Contract No. DE-AC02-06CH11357.


\clearpage
\begin{widetext}
\begin{center}
\textbf{\large Supplemental Material for ``Non-Hermitian phase transition from a polariton Bose-Einstein condensate to a photon laser''}
\end{center}
\end{widetext}
\setcounter{equation}{0}
\setcounter{figure}{0}
\setcounter{table}{0}
\setcounter{page}{1}


\author{Ryo Hanai}
\email{hanai@acty.phys.sci.osaka-u.ac.jp}
\affiliation{James Franck Institute and Department of Physics, University of Chicago, Illinois, 60637, USA} 
\affiliation{Department of Physics, Osaka University, Toyonaka 560-0043, Japan} 
\author{Alexander Edelman}
\affiliation{James Franck Institute and Department of Physics, University of Chicago, Illinois, 60637, USA} 
\author{Yoji Ohashi}
\affiliation{Department of Physics, Keio University, Yokohama 223-8522, Japan} 
\author{Peter B. Littlewood}
\affiliation{James Franck Institute and Department of Physics, University of Chicago, Illinois, 60637, USA} 
\affiliation{Physical Sciences and Engineering, Argonne National Laboratory, Argonne, Illinois 60439, USA}

\maketitle

\renewcommand{\thefigure}{S\arabic{figure}}
\renewcommand{\theequation}{S\arabic{equation}}

\section{Model}

We provide here the explicit form of the Hamiltonian $H$ of our model, depicted schematically in Fig. 2 in the main text \cite{Hanai2018_supp,Yamaguchi2012_supp,Yamaguchi2013_supp,Yamaguchi2015_supp}. 
The Hamiltonian is given by the sum of three parts $H=H_{\rm s}+H_{\rm env}+H_{\rm t}$. Here,
\begin{eqnarray}
&&H_{\rm s}
=
\sum_{\bm k,\sigma={\rm e,h}}\varepsilon_{\bm k,\sigma}
c^\dagger_{\bm k,\sigma}c_{\bm k,\sigma}
+\sum_{\bm q}\varepsilon_{\bm q}^{\rm cav}a^\dagger_{\bm q}a_{\bm q}
+\sum_{\bm k,\bm k',\bm q}
V_{\bm k-\bm k'}
\nonumber\\
&&\times\bigg[
\sum_{\sigma={\rm e,h}}
c^\dagger_{\bm k+\bm q/2,\sigma}c^\dagger_{-\bm k+\bm q/2,\sigma}c_{-\bm k'+\bm q/2,\sigma}c_{\bm k'+\bm q/2,\sigma}
\nonumber\\
&& \ \ 
-c^\dagger_{\bm k+\bm q/2,{\rm e}}c^\dagger_{-\bm k+\bm q/2,{\rm h}}c_{-\bm k'+\bm q/2,{\rm h}}c_{\bm k'+\bm q/2,{\rm e}}
\bigg]
\nonumber\\
&&+\sum_{\bm k,\bm q}[g c^\dagger_{\bm p+\bm q/2,{\rm e}}c^\dagger_{-\bm k+\bm q/2,{\rm h}}a_{\bm q}+{\rm h.c.}],
\end{eqnarray}
is the system Hamiltonian composed of electrons, holes, and photons.
$c_{\bm p,{\rm e(h)}}$ is an annihilation operator of an electron (hole) and
$\varepsilon_{\bm p,{\rm e(h)}}=\hbar^2\bm p^2/(2m_{{\rm e(h)}})+E_{\rm g}/2$ is the kinetic energy of an electron (hole), where $m_{\rm e(h)}$ is the effective mass of an electron (hole) in the conduction (valence) band
and $E_{\rm g}$ is the energy gap of the material.
$a_{\bm q}$ is an annihilation operator of a photon in the cavity, and $\varepsilon_{\bm q}^{\rm cav}=\hbar\omega_{\rm cav}+\hbar^2\bm q^2/(2m_{\rm cav})$ is the kinetic energy of photons, where $\hbar\omega_{\rm cav}=(c/n_{\rm c})\hbar(2\pi/\lambda)$ can be controlled by varying the microcavity length $\lambda$ ($n_{\rm c}$ is the refractive index of the microcavity).
The second term describes the pair-annihilation (creation) of electrons and holes accompanied by creation (annihilation) of photons, where $g$ is the dipole coupling constant.
The last term describes the repulsive and attractive Coulomb interactions between the electrons and holes, where $V_{\bm k-\bm k'}=e^2/(2\epsilon|\bm k-\bm k'|)$ 
($\epsilon$ is the dielectric constant).

Incoherent pumping of electrons and holes is modeled as a coupling to a (free) bath via the tunneling coefficient $\Gamma_{\rm b,{\rm e(h)}}$. Similarly, we model the photon decay as a coupling to a (free) vacuum via $\Gamma_{\rm v}$.
These are described by the Hamiltonian, 
\begin{eqnarray}
&&H_{\rm t}=\sum_{\bm k,\bm K,\sigma={\rm e,h},i}[\Gamma_{\rm b,\sigma}c^\dagger_{\bm k,\sigma}b_{\bm K,\sigma}e^{i\bm k\cdot\bm r_i}e^{-i\bm K\cdot \bm R_i}+{\rm h.c.}]
\nonumber\\
&&\ \ \ \ \ \ \ \ \ \ +
\sum_{\bm q,\bm Q,i}[\Gamma_{\rm v}
a^\dagger_{\bm q}\psi_{\bm Q}e^{i\bm q\cdot\bm r_i}e^{-i\bm Q\cdot \bm R_i}+{\rm h.c.}],
\\
&&H_{\rm env}
=\sum_{\bm P,\sigma={\rm e,h}}\varepsilon_{\bm P,\sigma}^{\rm b}b^\dagger_{\bm P,\sigma}b_{\bm P,\sigma}
+\sum_{\bm Q}\varepsilon_{\bm Q}^{\rm ph,v}\psi^\dagger_{\bm Q}\psi_{\bm Q}.
\end{eqnarray}
Here, 
$b_{\bm P,{\rm e(h)}}$ and $\psi_{\bm Q}$ are annihilation operators of the bath electrons (holes) and the vacuum photons, respectively, 
and $\varepsilon_{\bm P,{\rm e(h)}}^{\rm b}$ and $\varepsilon_{\bm Q}^{\rm v}$ are the kinetic energy of the bath electrons (holes) and the vacuum photons, respectively. 
We have assumed that the carriers tunnel from position $\bm r_i$ in the system to $\bm R_i$ in the bath or vacuum $(i=1,2,...,N_{\rm t})$. The positions $\bm r_i$ and $\bm R_i$ are assumed to be randomly distributed, in order to model homogeneous pumping and decay of carriers \cite{Hanai2018_supp}. 
As shown soon later, this results in a decay rate of photons given by
\begin{eqnarray}
\kappa = \pi N_{\rm t}|\Gamma_{\rm v}|^2 \rho_{\rm v},
\label{kappa}
\end{eqnarray}
and an incoherent pumping rate of the electrons (holes)
\begin{eqnarray}
\gamma_{\rm e(h)} = \pi N_{\rm t}|\Gamma_{\rm b,{\rm e(h)}}|^2 \rho_{\rm b,e(h)}.
\label{gamma_sig}
\end{eqnarray}
Here, the bath electron (hole) density of states $\rho_{\rm b,e(h)}$ and the vacuum photon density of states $\rho_{\rm v}$ are both assumed to be white 
(i.e., $\rho_{\rm v}={\rm const.}, \rho_{\rm b,\sigma={\rm e,h}}={\rm const.}$).

For the system to converge into a steady state, 
we assume that the bath and the vacuum are large compared to the system such that they stay in equilibrium.
The bath electron and hole distribution is given by the Fermi distribution function, 
\begin{eqnarray}
f_{{\rm b},\sigma={\rm e,h}}(\omega)
=\frac{1}{e^{(\hbar\omega-\mu_{\rm b,\sigma})/T_{\rm b}}+1},
\end{eqnarray}
characterized by the bath temperature $T_{\rm b}$ and the electron and hole chemical potential $\mu_{{\rm b},\sigma={\rm e,h}}$.
The vacuum photon distribution is given by, $f_{\rm v}(\omega)=0$. 
 
\section{Derivation of the equation of motion}

We now derive the general form of the equation of motion of the above model, which turns out to be given by the generalized Boltzmann equations [Eqs. (1) and (2) in the main text] and the Heisenberg equation of the photon amplitude [Eq. (3) in the main text].

Let us first derive the former. 
To study the dynamics of an interacting many-body system, it is convenient to consider the Nambu-Keldysh single-particle Green's function of electrons and holes, defined by \cite{Rammer_supp}, 
\begin{widetext}
\begin{eqnarray}
&&\hat G (\bm r_1,t_1;\bm r_2,t_2)
=\left(
\begin{array}{cc}
\hat G^{\rm R}(\bm r_1,t_1;\bm r_2,t_2) &
\hat G^{\rm K}(\bm r_1,t_1;\bm r_2,t_2) \\
0 &
\hat G^{\rm A}(\bm r_1,t_1;\bm r_2,t_2) 
\end{array}
\right)
\nonumber\\
&&
=-\frac{i}{\hbar}\left(
\begin{array}{cc}
\theta(t_1-t_2)
\avg{\{\hat\Psi(\bm r_1,t_1)
\diamondcomma \hat\Psi^\dagger(\bm r_2,t_2) \} } & 
\avg{\hat\Psi(\bm r_1,t_1)
\diamond \hat\Psi^\dagger(\bm r_2,t_2)
-\hat\Psi^\dagger(\bm r_2,t_2)
\diamond
\hat\Psi(\bm r_1,t_1)}
\\
0 & 
-\theta(t_2-t_1)
\avg{\{\hat\Psi(\bm r_1,t_1)
\diamondcomma 
\hat\Psi^\dagger(\bm r_2,t_2) \} }
\end{array}
\right),
\end{eqnarray}
where $\theta(x)$ is a step funtion. Here, we have introduced a Nambu operator
\begin{eqnarray}
\hat \Psi(\bm r,t)
=\left(
\begin{array}{c}
c_{\rm e}(\bm r,t) \\
c_{\rm h}^\dagger(\bm r,t)
\end{array}
\right)
\equiv\left(
\begin{array}{c}
\Psi_1(\bm r,t) \\
\Psi_2(\bm r,t)
\end{array}
\right),
\end{eqnarray}
and the product
\begin{eqnarray}
\big(
\hat \Psi(\bm r_1,t_1)\diamond \hat \Psi^\dagger(\bm r_2,t_2)
\big)_{s,s'}
&\equiv&
\Psi_{s}(\bm r_1,t_1)\Psi_{s'}^\dagger(\bm r_2,t_2)
=
\left(
\begin{array}{cc}
c_{\rm e}(\bm r_1,t_1)   c_{\rm e}^\dagger(\bm r_2,t_2)  & 
c_{\rm e}(\bm r_1,t_1)c_{\rm h}(\bm r_2,t_2)\\
c_{\rm h}^\dagger(\bm r_1,t_1) c_{\rm e}^\dagger(\bm r_2,t_2) & 
c_{\rm h}^\dagger(\bm r_1,t_1) c_{\rm h}(\bm r_2,t_2)
\end{array}
\right)_{s,s'}
,\\
\big(
\hat \Psi^\dagger(\bm r_2,t_2)\diamond \hat \Psi(\bm r_1,t_1)
\big)_{s,s'}
&\equiv&
\Psi^\dagger_{s'}(\bm r_2,t_2)\Psi_{s}(\bm r_1,t_1)
=\left(
\begin{array}{cc}
c^\dagger_{\rm e}(\bm r_2,t_2) c_{\rm e}(\bm r_1,t_1)              & 
c_{\rm h}(\bm r_2,t_2) c_{\rm e}(\bm r_1,t_1)\\
c_{\rm e}^\dagger(\bm r_2,t_2) c_{\rm h}^\dagger(\bm r_1,t_1) & c_{\rm h}(\bm r_2,t_2) c_{\rm h}^\dagger(\bm r_1,t_1) 
\end{array}
\right)_{s,s'}.
\end{eqnarray}
An especially important quantity of interest is the lesser Green's function,
\begin{eqnarray}
\hat G^{\rm <}(\bm r_1,t_1;\bm r_2,t_2)
&=&\frac{1}{2}[-\hat G^{\rm R}+\hat G^{\rm A}+\hat G^{\rm K}](\bm r_1,t_1;\bm r_2,t_2)
=\frac{i}{\hbar}
\avg{\hat\Psi^\dagger(\bm r_2,t_2)\diamond\hat\Psi(\bm r_1,t_1)}
\nonumber\\
&=&
\frac{i}{\hbar}\left(
\begin{array}{cc}
\avg{c^\dagger_{\rm e}(\bm r_2,t_2) c_{\rm e}(\bm r_1,t_1)}              & 
\avg{c_{\rm h}(\bm r_2,t_2) c_{\rm e}(\bm r_1,t_1)} \\
\avg{c_{\rm e}^\dagger(\bm r_2,t_2) c_{\rm h}^\dagger(\bm r_1,t_1)} & 
\avg{c_{\rm h}(\bm r_2,t_2) c_{\rm h}^\dagger(\bm r_1,t_1) }
\end{array}
\right),
\end{eqnarray}
which directly relates to the electron (hole) density $n_{\bm k,{\rm e(h)}}(\bm r,t)$ and the polarization $p_{\bm k}(\bm r,t)$.
By transforming this quantity to the so-called Wigner representation, where the coordinates $(\bm r_1,t_1)$ and $(\bm r_2,t_2)$ are rewritten in terms of the relative coordinate $\bm r_{\rm r}=\bm r_1-\bm r_2,t_{\rm r}=t_1-t_2$ and the center of motion coordinate 
$\bm r=(\bm r_1+\bm r_2)/2,t=(t_1+t_2)/2$, 
the electron (hole) density $n_{\bm k,\sigma={\rm e(h)}}(\bm r,t)$ 
and the electron-hole dipole polarization $p_{\bm k}(\bm r,t)$ are obtained as,
\begin{eqnarray}
\left(
\begin{array}{cc}
n_{\bm k,{\rm e}}(\bm r,t) & p_{\bm k}(\bm r,t)\\
p_{\bm k}^*(\bm r,t) & 1-n_{\bm k,{\rm h}}(\bm r,t)
\end{array}
\right)
&=&-i \hbar
\int d\bm r_{\rm r} e^{-i\bm k\cdot \bm r_{\rm r}} 
\hat G^<(\bm r_{\rm r},t_{\rm r}=0;\bm r,t)
=-i\hbar \int_{-\infty}^{\infty}\frac{d\omega}{2\pi} 
\hat G^<(\bm k,\omega;\bm r,t).
\end{eqnarray}
Below, we show that the equation of motion of these valuables are given by the generalized Boltzmann equations (1) and (2).

The dynamics of the single-particle Green's function $\hat G$ is determined by the Dyson's equation \cite{Rammer_supp},
\begin{eqnarray}
\hat G=\hat G_0 
+ \hat G_0 \otimes \hat \Sigma \otimes \hat G,
\label{Dyson}
\end{eqnarray}
where we have introduced a short-hand notation,
\begin{eqnarray}
[\hat A \otimes \hat B](\bm r_1,t_1;\bm r_2,t_2)
=\int_{-\infty}^{\infty} dt_1'\int d\bm r_1' 
\hat A(\bm r_1,t_1;\bm r_1',t_1')\hat B(\bm r_1',t_1';\bm r_2,t_2),
\end{eqnarray}
and have omitted the space-time index in Eq. (\ref{Dyson}).
The (Fourier transformed) free electron-hole Green's function is given by,
\begin{eqnarray}
&&\hat G_0(\bm k,\omega)
=
\left(
\begin{array}{cc}
\hat G^{\rm R}_0(\bm k,\omega) & \hat G^{\rm K}_0(\bm k,\omega) \\
0                                         & \hat G^{\rm A}_0(\bm k,\omega)
\end{array}
\right)
\end{eqnarray}
with
\begin{eqnarray}
G_0^{\rm R}(\bm k,\omega)&=&
\left(
\begin{array}{cc}
\hbar\omega+i\delta - \varepsilon_{\bm k,{\rm e}} & 0 \\
0 & \hbar\omega+i\delta + \varepsilon_{\bm k,{\rm h}} 
\end{array}
\right)^{-1},\\
G_0^{\rm A}(\bm k,\omega)&=&
\left(
\begin{array}{cc}
\hbar\omega-i\delta - \varepsilon_{\bm k,{\rm e}} & 0 \\
0 & \hbar\omega-i\delta + \varepsilon_{\bm k,{\rm h}} 
\end{array}
\right)^{-1} , \\
G_0^{\rm K}(\bm k,\omega)&=&
\left(
\begin{array}{cc}
-2\pi i [1-2f(\omega)]\delta(\hbar\omega-\varepsilon_{\bm k,{\rm e}}) & 0 \\
0 & 2\pi i [1-2f(-\omega)]\delta(\hbar\omega+\varepsilon_{\bm k,{\rm h}})
\end{array}
\right), 
\end{eqnarray}
where $\tau_{i=1,2,3}$ are Pauli matrices acting on the Nambu space.
Here, $f(\omega)$ in the Keldysh component is the initial distribution of the relevant system, which, as shown below, does not affect the final form of the equation of motion.
The effects of the many-body interaction and the coupling to the bath are described by the self-energy,
\begin{eqnarray}
\hat \Sigma(\bm r_1,t_1;\bm r_2,t_2)
=
\left(
\begin{array}{cc}
\hat \Sigma^{\rm R}(\bm r_1,t_1;\bm r_2,t_2) & 
\hat \Sigma^{\rm K}(\bm r_1,t_1;\bm r_2,t_2) \\
0                                                      & 
\hat\Sigma^{\rm A}(\bm r_1,t_1;\bm r_2,t_2)
\end{array}
\right).
\end{eqnarray}
We can proceed by formally solving the Dyson's equation (\ref{Dyson}) as,
\begin{eqnarray}
\hat G^{\rm R}
&=&\big[ [G_0^{\rm R}]^{-1} - \hat \Sigma^{\rm R}\big]^{-1},
\label{GR}
\\
\hat G^{\rm A}
&=&\big[ [G_0^{\rm A}]^{-1} - \hat \Sigma^{\rm A}\big]^{-1}
=[\hat G^{\rm R}]^\dagger,
\label{GA}
\\
\hat G^{\rm K}
&=&\hat G^{\rm R} \otimes \hat \Sigma^{\rm K}\otimes \hat G^{\rm A}
+(1+\hat G^{\rm R}\otimes\hat \Sigma^{\rm R})
\hat G^{\rm K}_0(1+\hat \Sigma^{\rm A}\otimes \hat G^{\rm A})
\nonumber\\
&=&
\hat G^{\rm R} \otimes \hat \Sigma^{\rm K}\otimes \hat G^{\rm A}
+\hat G^{\rm R}\otimes [\hat G^{\rm R}_0]^{-1}
\otimes \hat G^{\rm K}_0 \otimes [\hat G^{\rm A}_0]^{-1} \otimes \hat G^{\rm A}
\nonumber\\
&=&
\hat G^{\rm R} \otimes \hat \Sigma^{\rm K}\otimes \hat G^{\rm A}.
\label{GK}
\end{eqnarray}
In deriving Eq. (\ref{GK}), we have used Eqs. (\ref{GR}) and (\ref{GA}) in the second equality and have used the relation, 
\begin{eqnarray}
&&\big[
[\hat G^{\rm R}_0]^{-1}
\otimes \hat G^{\rm K}_0 \otimes [\hat G^{\rm A}_0]^{-1}
\big](\bm k,\omega)
=
[\hat G^{\rm R}_0]^{-1}(\bm k,\omega)
\hat G^{\rm K}_0(\bm k,\omega) 
[\hat G^{\rm A}_0]^{-1}(\bm k,\omega)
=0,
\end{eqnarray}
in the third.
From Eq. (\ref{GK}), the lesser Green's function $\hat G^<$ satisfies,
\begin{eqnarray}
0&=&
[\hat G^{\rm R}-\hat G^{\rm A}+\hat G^<]
-\hat G^{\rm R}
\otimes
[\hat \Sigma^{\rm R}-\hat \Sigma^{\rm A}+\hat \Sigma^<]
\otimes
\hat G^{\rm A}
\nonumber\\
&=&\hat G^{\rm R}\otimes[1+\hat \Sigma^{\rm A}\otimes G^{\rm A}]
-[1+\hat G^{\rm R}\otimes \hat \Sigma^{\rm R}]\otimes\hat G^{\rm A}
+\hat G^< 
-\hat G^{\rm R}\otimes \hat \Sigma^< \otimes G^{\rm A} 
\nonumber\\
&=&
\Big[\hat G^{\rm R}\otimes[\hat G^{\rm A}_0]^{-1}\otimes\hat G^{\rm A}
-\hat G^{\rm R}\otimes[\hat G^{\rm R}_0]^{-1}\otimes\hat G^{\rm A}
\Big]
+\Big[\hat G^<
-\hat G^{\rm R}\otimes\Sigma^< \otimes\hat G^{\rm A}
\Big]
=\hat G^<
-\hat G^{\rm R}\otimes\hat \Sigma^< \otimes\hat G^{\rm A},
\label{G<derive}
\end{eqnarray}
or
\begin{eqnarray}
\hat G^{\rm <}
=\hat G^{\rm R} \otimes \hat \Sigma^{\rm <}\otimes \hat G^{\rm A},
\label{G<}
\end{eqnarray}
where 
\begin{eqnarray}
\hat \Sigma^<=\frac{1}{2}[-\hat \Sigma^{\rm R}+\hat \Sigma^{\rm A}+\hat \Sigma^{\rm K}],
\label{Sigma<}
\end{eqnarray}
is the lesser component of the self-energy.
We have used Eqs. (\ref{GR}) and (\ref{GA}) in obtaining the third equality of  Eq. (\ref{G<derive}) and 
$[\hat G^{\rm R}_0]^{-1}=[\hat G^{\rm A}_0]^{-1}$ in the last.
This yields,
\begin{eqnarray}
[\hat G^{\rm R}_0]^{-1} \otimes \hat G^{\rm <}
&=&
\hat \Sigma^{\rm R}\otimes \hat G^{\rm <}
+\hat \Sigma^{\rm <}\otimes \hat G^{\rm A},
\label{G<_r}
\\
\hat G^{\rm <}\otimes [\hat G^{\rm A}_0]^{-1}
&=&
\hat G^{\rm <}\otimes \hat \Sigma^{\rm A}
+\hat G^{\rm R}\otimes \hat \Sigma^{\rm <},
\label{G<_l}
\end{eqnarray}
giving,
\begin{eqnarray}
-[\hat G^{\rm R}_0]^{-1} \otimes \hat G^{\rm <}
+\hat G^{\rm <}\otimes [\hat G^{\rm A}_0]^{-1}
&=&
-\hat \Sigma^{\rm R}\otimes \hat G^{\rm <}
+\hat G^{\rm <}\otimes \hat \Sigma^{\rm A}
-\hat \Sigma^{\rm <}\otimes \hat G^{\rm A}
+\hat G^{\rm R}\otimes \hat \Sigma^{\rm <}.
\label{G<_rl}
\end{eqnarray}

Let us obtain the explicit form of the left-hand side of Eq. (\ref{G<_rl}). 
The two terms on the left-hand side is written as,
\begin{eqnarray}
\big[
[\hat G^{\rm R}_0]^{-1}\otimes \hat G^{\rm <}
\big]
(\bm r_1,t_1;\bm r_2,t_2)
&=&\left(
\begin{array}{cc}
i\hbar\frac{\overrightarrow{\partial}}{\partial t_1} 
- \big( -\frac{\hbar^2\overrightarrow{\nabla}_1^2}{2m_{\rm e}}
+ \frac{E_{\rm g}}{2}\big)
& 0 \\ 
0 
& i\hbar\frac{\overrightarrow{\partial}}{\partial t_1}
+ \big( -\frac{\hbar^2\overrightarrow{\nabla}_1^2}{2m_{\rm h}}
+\frac{E_{\rm g}}{2}\big)
\end{array}
\right)
\hat G^{\rm <}
(\bm r_1,t_1;\bm r_2,t_2), 
\label{G0R-1G<}
\\
\big[
\hat G^{\rm <}\otimes [\hat G^{\rm A}_0]^{-1}
\big]
(\bm r_1,t_1;\bm r_2,t_2)
&=&
\hat G^{\rm <}
(\bm r_1,t_1;\bm r_2,t_2)
\left(
\begin{array}{cc}
i\hbar\frac{\overleftarrow{\partial}}{\partial t_2} 
- \big( -\frac{\hbar^2\overleftarrow{\nabla}_2^2}{2m_{\rm e}}
+ \frac{E_{\rm g}}{2}\big)
& 0 \\ 
0 
& i\hbar\frac{\overleftarrow{\partial}}{\partial t_2}
+ \big( -\frac{\hbar^2\overleftarrow{\nabla}_2^2}{2m_{\rm h}}
+\frac{E_{\rm g}}{2}\big)
\end{array}
\right),
\label{G<G0A-1}
\end{eqnarray}
where the partial derivatives with arrows pointing to the right (left) operates to the quantity on the right (left).
In the Wigner representation, Eqs. (\ref{G0R-1G<}) and (\ref{G<G0A-1}) are expressed as,
\begin{eqnarray}
&&\big[
[\hat G^{\rm R}_0]^{-1}\otimes \hat G^{\rm <}
\big]
(\bm k,\omega;\bm r,t)
\nonumber\\
&&
=\left(
\begin{array}{cc}
\frac{i\hbar}{2}\frac{\overrightarrow{\partial}}{\partial t} 
+\hbar\omega
- \Big[ -\frac{\hbar^2}{2m_{\rm e}}
\big(\frac{\overrightarrow{\nabla}}{2}+i\bm k\big)^2
+ \frac{E_{\rm g}}{2}\Big]
& 0 \\ 
0 
& \frac{i\hbar}{2}\frac{\overrightarrow{\partial}}{\partial t} 
+\hbar\omega
+ \Big[ -\frac{\hbar^2}{2m_{\rm h}}
\big(\frac{\overrightarrow{\nabla}}{2}+i\bm k\big)^2
+\frac{E_{\rm g}}{2}\Big]
\end{array}
\right)
\hat G^{\rm <}
(\bm k,\omega;\bm r,t), 
\\
&&\big[
\hat G^{\rm <}\otimes [\hat G^{\rm A}_0]^{-1}
\big]
(\bm k,\omega;\bm r,t)
\nonumber\\
&&
=
\hat G^{\rm <}
(\bm k,\omega;\bm r,t)
\left(
\begin{array}{cc}
-\frac{i\hbar}{2}\frac{\overleftarrow{\partial}}{\partial t} 
+\hbar\omega
- \Big[ -\frac{\hbar^2}{2m_{\rm e}}
\big(\frac{\overleftarrow{\nabla}}{2}-i\bm k\big)^2
+ \frac{E_{\rm g}}{2}\Big]
& 0 \\ 
0 
& -\frac{i\hbar}{2}\frac{\overleftarrow{\partial}}{\partial t} 
+\hbar\omega 
+ \big[ -\frac{\hbar^2}{2m_{\rm h}}
\Big(\frac{\overleftarrow{\nabla}}{2}-i\bm k\big)^2
+\frac{E_{\rm g}}{2}\Big]
\end{array}
\right).
\end{eqnarray}

Integrating both sides of Eq. (\ref{G<_rl}) over $\omega$, we obtain the generalized Boltzmann equation,
\begin{eqnarray}
&&\left(
\begin{array}{cc}
\frac{\partial}{\partial t}
n_{\bm k,{\rm e}}(\bm r,t)
+\bm v_{\bm k,{\rm e}}\cdot \nabla
n_{\bm k,{\rm e}}(\bm r,t)
& \frac{\partial}{\partial t}
p_{\bm k}(\bm r,t)
+\frac{i}{\hbar}
\big(
\varepsilon_{\bm k,{\rm e}}+\varepsilon_{\bm k,{\rm h}}
-\frac{\hbar^2\nabla^2}{4m_{\rm eh}}
\big)
p_{\bm k}(\bm r,t)
\\
\frac{\partial}{\partial t}
p_{\bm k}^*(\bm r,t)
-\frac{i}{\hbar}\big(
\varepsilon_{\bm k,{\rm e}}+\varepsilon_{\bm k,{\rm h}}
-\frac{\hbar^2\nabla^2}{4m_{\rm eh}}
\big)
p_{\bm k}^*(\bm r,t)
& -\frac{\partial}{\partial t}
n_{\bm k,{\rm h}}(\bm r,t)
-\bm v_{\bm k,{\rm h}}\cdot \nabla
n_{\bm k,{\rm h}}(\bm r,t)
\end{array}
\right)
\nonumber\\
&&=
\int_{-\infty}^{\infty} \frac{d\omega}{2\pi}
\big[
-\hat \Sigma^{\rm R}\otimes \hat G^{\rm <}
+\hat G^{\rm <}\otimes \hat \Sigma^{\rm A}
-\hat \Sigma^{\rm <}\otimes \hat G^{\rm A}
+\hat G^{\rm R}\otimes \hat \Sigma^{\rm <}
\big](\bm k,\omega;\bm r,t),
\end{eqnarray}
where $2 m_{\rm eh}^{-1}=m_{\rm e}^{-1}+m_{\rm h}^{-1}$ is twice the reduced mass. 
The right-hand side can be interpreted as the collision term.
Note that, unlike in the conventional Boltzmann equation, the collision term depends explicitly on time and space.

\begin{figure}
\begin{center}
\includegraphics[width=0.25\linewidth,keepaspectratio]{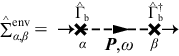}
\end{center}
\caption{(Color online) Diagramatic expression $\hat\Sigma_{\rm env}$. The dashed line represents the bath Green's function $\hat B_{\rm b}$ and the cross represents $\hat \Gamma_{\rm b}$.
}
\label{fig_env}
\end{figure}

To show that the coupling to the bath induces dephasing and decay, we separate the self-energy into two terms, 
\begin{eqnarray}
\hat\Sigma 
= \hat\Sigma_{\rm env}+\hat\Sigma_{\rm int},
\end{eqnarray}
where the first term ($\hat\Sigma_{\rm env}$) describes the effects from the system-bath coupling and the second term ($\hat \Sigma_{\rm int}$) describes the many-body interaction effects.
We note that the cross-term $\hat\Sigma_{\rm env-int}$ is absent since we have assumed that the bath is large compared to the system.
The diagrammatic expression of $\hat\Sigma_{\rm env}$ is shown in Fig. \ref{fig_env}, where its explicit form is given by,
\begin{eqnarray}
\hat\Sigma_{\rm env}^{\rm R}(\bm k,\omega;\bm r,t)
=\hat\Sigma_{\rm env}^{\rm R} 
&=& N_{\rm t}\sum_{\bm P}
\hat \Gamma_{\rm b}^\dagger \hat B_{\rm b}^{\rm R}(\bm P,\omega) \hat \Gamma_{\rm b}
=\left(
\begin{array}{cc}
-i\gamma_{\rm e} & 0 \\
0 & -i\gamma_{\rm h} 
\end{array}
\right), 
\label{Sigma_envR}
\\
\hat\Sigma_{\rm env}^{\rm A}(\bm k,\omega;\bm r,t)
=\hat\Sigma_{\rm env}^{\rm A} 
&=& N_{\rm t}\sum_{\bm P}
\hat \Gamma_{\rm b}^\dagger \hat B_{\rm b}^{\rm A}(\bm P,\omega) \hat \Gamma_{\rm b}
=\left(
\begin{array}{cc}
i\gamma_{\rm e} & 0 \\
0 & i\gamma_{\rm h} 
\end{array}
\right), 
\label{Sigma_envA}
\\
\hat\Sigma_{\rm env}^{\rm K}(\bm k,\omega;\bm r,t)
=\hat\Sigma_{\rm env}^{\rm K}(\omega) 
&=& N_{\rm t}\sum_{\bm P}
\hat \Gamma_{\rm b}^\dagger \hat B_{\rm b}^{\rm K}(\bm P,\omega) \hat \Gamma_{\rm b}
=\left(
\begin{array}{cc}
 2i\gamma_{\rm e}[1-2f_{\rm b,{\rm e}}(\omega)] & 0 \\
0 & -2i\gamma_{\rm h} [1-2f_{\rm b,{\rm h}}(-\omega)]
\end{array}
\right), 
\end{eqnarray}
and the lesser component is given by,
\begin{eqnarray}
\hat\Sigma_{\rm env}^<(\omega)
=2i\left(
\begin{array}{cc}
\gamma_{\rm e} f_{\rm b,{\rm e}}(\omega) & 0 \\
0 & \gamma_{\rm h}f_{\rm b,{\rm h}}(-\omega)
\end{array}
\right).
\label{Sigma_env<}
\end{eqnarray}
Here, $\hat\Gamma_{\rm b}
={\rm diag}(\Gamma_{\rm b,{\rm e}},\Gamma_{\rm b,{\rm h}})$ and 
\begin{eqnarray}
\hat B_{\rm b}^{\rm R}(\bm k,\omega)&=&
\left(
\begin{array}{cc}
\hbar\omega+i\delta - \varepsilon_{\bm k,{\rm e}}^{\rm b} & 0 \\
0 & \hbar\omega+i\delta + \varepsilon_{\bm k,{\rm h}}^{\rm b} 
\end{array}
\right)^{-1},
\label{BbR}
\\
\hat B_{\rm b}^{\rm A}(\bm k,\omega)&=&
\left(
\begin{array}{cc}
\hbar\omega-i\delta - \varepsilon_{\bm k,{\rm e}}^{\rm b} & 0 \\
0 & \hbar\omega-i\delta + \varepsilon_{\bm k,{\rm h}}^{\rm b} 
\end{array}
\right)^{-1} , 
\label{BbA}
\\
\hat B_{\rm b}^{\rm K}(\bm k,\omega)&=&
\left(
\begin{array}{cc}
-2\pi i [1-2f_{\rm b,{\rm e}}(\omega)]\delta(\hbar\omega-\varepsilon_{\bm k,{\rm e}}^{\rm b}) & 0 \\
0 & 2\pi i [1-2f_{\rm b,{\rm h}}(-\omega)]\delta(\hbar\omega+\varepsilon_{\bm k,{\rm h}}^{\rm b})
\end{array}
\right), 
\label{BbK}
\end{eqnarray}
is the electron-hole single-particle Green's function in the bath.
In the derivation, we have assumed that the bath is white 
($\rho_{\rm b,e(h)}=i\sum_{\bm P}[B_{\rm b}^{\rm R}]_{11(22)}(\bm P,\omega)={\rm const.}$) and used the definition of $\gamma_\sigma$ given by Eq. (\ref{gamma_sig}).
Since we have assumed that the bath is large compared to the system, the bath Green's function is unaffected by the system dynamics.
From Eqs. (\ref{Sigma_envR})-(\ref{Sigma_env<}), 
\begin{eqnarray}
&&
\int_{-\infty}^{\infty}\frac{d\omega}{2\pi}
[
-\hat \Sigma^{\rm R}_{\rm env}\otimes \hat G^{\rm <}
+\hat G^{\rm <}\otimes \hat \Sigma^{\rm A}_{\rm env}
-\hat \Sigma^{\rm <}_{\rm env}\otimes \hat G^{\rm A}
+\hat G^{\rm R}\otimes \hat \Sigma^{\rm <}_{\rm env}
]
(\bm k,\omega;\bm r,t)
\nonumber\\
&&=
-\frac{1}{\hbar}
\left(
\begin{array}{cc}
2\gamma_{\rm e}[n_{\bm k,{\rm e}}(\bm r,t) - n_{\bm k,{\rm e}}^{\rm env}(\bm r,t)] & 
2\gamma[p_{\bm k}(\bm r,t) - p_{\bm k}^{\rm env}(\bm r,t)] \\
2\gamma[p_{\bm k}^*(\bm r,t)- p_{\bm k}^{\rm env*}(\bm r,t)] & 
-2\gamma_{\rm h}[n_{\bm k,{\rm h}}(\bm r,t) - n_{\bm k,{\rm h}}^{\rm env}(\bm r,t)]
\end{array}
\right),
\end{eqnarray}
where $\gamma=(\gamma_{\rm e}+\gamma_{\rm h})/2$ and
\begin{eqnarray}
n_{\bm k,{\rm e}}^{\rm env}(\bm r,t)
&=& 
\frac{\hbar}{2\gamma_{\rm e}}
\int_{-\infty}^{\infty} \frac{d\omega}{2\pi}
[
-\hat \Sigma^{\rm <}_{\rm env}\otimes \hat G^{\rm A}
+\hat G^{\rm R}\otimes \hat \Sigma^{\rm <}_{\rm env}
]_{11}
(\bm k,\omega;\bm r,t), \\
1-n_{\bm k,{\rm h}}^{\rm env} (\bm r,t)
&=&
\frac{\hbar}{2\gamma_{\rm h}}
\int_{-\infty}^{\infty} \frac{d\omega}{2\pi}
[
-\hat \Sigma^{\rm <}_{\rm env}\otimes \hat G^{\rm A}
+\hat G^{\rm R}\otimes \hat \Sigma^{\rm <}_{\rm env}
]_{22}
(\bm k,\omega;\bm r,t), \\
p_{\bm k}^{\rm env}(\bm r,t)
&=&
\frac{\hbar}{2\gamma}
\int_{-\infty}^{\infty} \frac{d\omega}{2\pi}
[
-\hat \Sigma^{\rm <}_{\rm env}\otimes \hat G^{\rm A}
+\hat G^{\rm R}\otimes \hat \Sigma^{\rm <}_{\rm env}
]_{12}
(\bm k,\omega;\bm r,t).
\end{eqnarray}
This gives
\begin{eqnarray}
&&\partial_t p_{\bm k}(\bm r,t)
=-
\frac{i}{\hbar}\Big(
\varepsilon_{\bm k,{\rm e}}
+\varepsilon_{\bm k,{\rm h}}
-\frac{\hbar^2\nabla^2}{4m_{\rm eh}}
-2i\gamma
\Big)
p_{\bm k}(\bm r,t)
+ I^{\rm pol}_{\bm k}(\bm r,t),
\label{pk_Boltzmann}
\\
&&\partial_t n_{\bm k,\sigma={\rm e,h}}(\bm r,t)
+\bm v_{\bm k,\sigma={\rm e,h}}\cdot \nabla n_{\bm k,\sigma={\rm e,h}}(\bm r,t)
=
-\frac{2\gamma_\sigma}{\hbar}  n_{\bm k,\sigma={\rm e,h}}(\bm r,t)
+I_{\bm k,\sigma={\rm e,h}}(\bm r,t),
\label{nk_Boltzmann}
\end{eqnarray}
where $\bm v_{\bm k,\sigma}=\hbar\bm k/m_\sigma$, and 
\begin{eqnarray}
I_{\bm k,{\rm e}}(\bm r,t)
&=&\frac{2\gamma_{\rm e}}{\hbar} n_{\bm k,{\rm e}}^{\rm env}(\bm r,t)
+
\int_{-\infty}^{\infty}\frac{d\omega}{2\pi}
\big[
-\hat \Sigma^{\rm R}_{\rm int}\otimes \hat G^{\rm <}
+\hat G^{\rm <}\otimes \hat \Sigma^{\rm A}_{\rm int}
-\hat \Sigma^{\rm <}_{\rm int}\otimes \hat G^{\rm A}
+\hat G^{\rm R}\otimes \hat \Sigma^{\rm <}_{\rm int}
\big]_{11}
(\bm k,\omega;\bm r,t),
\\
I_{\bm k,{\rm h}}(\bm r,t)
&=&\frac{2\gamma_{\rm h}}{\hbar} n_{\bm k,{\rm h}}^{\rm env}(\bm r,t)
-
\int_{-\infty}^{\infty}\frac{d\omega}{2\pi}
\big[
-\hat \Sigma^{\rm R}_{\rm int}\otimes \hat G^{\rm <}
+\hat G^{\rm <}\otimes \hat \Sigma^{\rm A}_{\rm int}
-\hat \Sigma^{\rm <}_{\rm int}\otimes \hat G^{\rm A}
+\hat G^{\rm R}\otimes \hat \Sigma^{\rm <}_{\rm int}
\big]_{22}
(\bm k,\omega;\bm r,t),
\\
I_{\bm k}^{\rm pol}(\bm r,t)
&=&
\frac{2\gamma}{\hbar} p_{\bm k}^{\rm env}(\bm r,t)
+
\int_{-\infty}^{\infty}\frac{d\omega}{2\pi}
\big[
-\hat \Sigma^{\rm R}_{\rm int}\otimes \hat G^{\rm <}
+\hat G^{\rm <}\otimes \hat \Sigma^{\rm A}_{\rm int}
-\hat \Sigma^{\rm <}_{\rm int}\otimes \hat G^{\rm A}
+\hat G^{\rm R}\otimes \hat \Sigma^{\rm <}_{\rm int}
\big]_{12}
(\bm k,\omega;\bm r,t).
\end{eqnarray}
Equation (\ref{nk_Boltzmann}) is the desired Boltzmann equation (2) for $n_{\bm k,\sigma}(\bm r,t)$.

Note that, the term $I_{\bm k}^{\rm pol}(\bm r,t)$ in Eq. (\ref{pk_Boltzmann}) should vanish in the normal phase, since $p_{\bm k}(\bm r,t)=\lambda_{\rm cav}(\bm r,t)=0$ in this phase, while 
the condensed phase is characterized by the order parameter (See Refs. \cite{Hanai2018_supp,Yamaguchi2012_supp,Yamaguchi2013_supp,Yamaguchi2015_supp} and the later discussion for the analysis within the Hartree-Fock-Bogoliubov approximation as an example.), 
\begin{eqnarray}
\Delta_{\bm k}(\bm r,t)
=\sum_{\bm k'}V_{\bm k-\bm k'}p_{\bm k'}(\bm r,t) - g\lambda_{\rm cav}(\bm r,t).
\end{eqnarray}
Since $\Delta_{\bm k}(\bm r,t)=0$ in the normal phase, $I^{\rm pol}_{\bm k}(\bm r,t)$ can be written in the form,
\begin{eqnarray}
I^{\rm pol}_{\bm k}(\bm r,t)
=\frac{i}{\hbar}\sum_{\bm k'}L_{\bm k,\bm k'}(\bm r,t)
\Delta_{\bm k'}(\bm r,t),
\end{eqnarray}
which gives our final form of the Boltzmann equation for $p_{\bm k}(\bm r,t)$ [Eq. (2) in the main text].

\begin{figure}
\begin{center}
\includegraphics[width=0.25\linewidth,keepaspectratio]{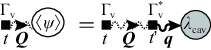}
\end{center}
\caption{(Color online) Diagramatic expression Eq. (\ref{psiQ}). The dotted curved line represents the vacuum photon Green's function $B_{\rm v}^{\rm R}$ and the solid square represents the tunneling $\Gamma_{\rm v}$.
}
\label{fig_psiQ}
\end{figure}

The other piece of interest is the dynamics of the photon amplitude $\lambda_{\rm cav}(\bm r,t)=\avg{a(\bm r,t)}$, given by Eq. (3) in the main text.
The Heisenberg equation of the photon annihilation operator $a(\bm r,t)$ is given by,
\begin{eqnarray}
i\hbar\partial_t a(\bm r,t)
=[a(\bm r,t),H]
&=&\Big(
\hbar\omega_{\rm cav}-\frac{\hbar^2\nabla^2}{2m_{\rm cav}}
\Big)a(\bm r,t)
+g\sum_{\bm k,\bm q}
e^{i\bm q\cdot \bm r}
c_{-\bm k+\bm q/2,{\rm h}}(t)c_{\bm k+\bm q/2,{\rm e}}(t)
\nonumber\\
&+&\sum_{\bm q,\bm Q,i}\Gamma_{\rm v}\psi_{\bm Q}(t)
e^{i\bm q  \cdot (\bm r-\bm r_i)}e^{-i\bm Q\cdot \bm R_i}.
\label{EOM_ph}
\end{eqnarray}
Taking the statistical average of Eq. (\ref{EOM_ph}), we get,
\begin{eqnarray}
&&i\hbar\partial_t \lambda_{\rm cav}(\bm r,t)
=\Big(
\hbar\omega_{\rm cav}-\frac{\hbar^2\nabla^2}{2m_{\rm cav}}
\Big)\lambda_{\rm cav}(\bm r,t)
+g\sum_{\bm k}
p_{\bm k}(\bm r,t)
+\avg{\sum_{\bm q,\bm Q,i}\Gamma_{\rm v}
\psi_{\bm Q}(t)
e^{i\bm q  \cdot (\bm r-\bm r_i)}e^{-i\bm Q\cdot \bm R_i}}.
\end{eqnarray}
By applying the Wick's theorem, as diagramatically described in Fig. \ref{fig_psiQ}, we obtain
\begin{eqnarray}
&&\avg{\sum_{\bm q,\bm Q,i}\Gamma_{\rm v}
\psi_{\bm Q}(t)
e^{i\bm q  \cdot (\bm r-\bm r_i)}e^{-i\bm Q\cdot \bm R_i}}
=N_{\rm t}|\Gamma_{\rm v}|^2
\int_{-\infty}^{\infty} dt'
\sum_{\bm Q}B^{\rm R}_{\rm v}(\bm Q,t-t')
\sum_{\bm q}e^{i\bm q\cdot \bm r} \avg{a_{\bm q}(t')}
\nonumber\\
&&
=N_{\rm t}|\Gamma_{\rm v}|^2
\int_{-\infty}^{\infty} dt'
\int_{-\infty}^{\infty} \frac{d\omega}{2\pi} e^{-i\omega (t-t')}
\sum_{\bm Q}\hat B^{\rm R}_{\rm v}(\bm Q,\omega)
\sum_{\bm q}e^{i\bm q\cdot \bm r} \avg{a_{\bm q}(t')}
=-i\kappa 
\int_{-\infty}^{\infty} dt'
\int_{-\infty}^{\infty} \frac{d\omega}{2\pi} e^{-i\omega (t-t')}
\sum_{\bm q}e^{i\bm q\cdot \bm r} \avg{a_{\bm q}(t')}
\nonumber\\
&&
=-i\kappa \lambda_{\rm cav}(\bm r,t),
\label{psiQ}
\end{eqnarray}
where 
$B_{\rm v}^{\rm R}(\bm Q,\omega)=[\hbar\omega-\varepsilon_{\bm Q}^{\rm v}+i\delta]^{-1}$
is the vacuum photon Green's function, and a white noise vacuum is assumed, i.e., 
$\rho_{\rm v}\equiv i\sum_{\bm Q}\hat B^{\rm R}_{\rm v}(\bm Q,\omega)/\pi={\rm const.}$ and the photon decay rate $\kappa$ is defined in Eq. (\ref{kappa}).
This yields the desired Heisenberg equation [Eq. (3) in the main text],
\begin{eqnarray}
i\hbar\partial_t\lambda_{\rm cav}(\bm r,t)
&=&
\Big[
\hbar\omega_{\rm cav}
-\frac{\hbar^2\nabla^2}{2m_{\rm cav}}
-i\kappa
\Big]
\lambda_{\rm cav}(\bm r,t) 
+g \sum_{\bm k}p_{\bm k}(\bm r,t).
\end{eqnarray}

\end{widetext}

\section{Proof of the existence of a phase boundary with an end point}

In the main text, we have shown from Eqs. (1) and (3) that, Eq. (4) in the main text,
\begin{eqnarray}
\hat A 
\left(
\begin{array}{c}
\lambda_{\rm cav}^0 \\
\lambda_{\rm eh}^0
\end{array}
\right)
=
\left(
\begin{array}{cc}
h_{\rm cav}  & g_0  \\
\tilde g_0^* & h_{\rm eh} 
\end{array}
\right)
\left(
\begin{array}{c}
\lambda_{\rm cav}^0 \\
\lambda_{\rm eh}^0
\end{array}
\right)
=E\left(
\begin{array}{c}
\lambda_{\rm cav}^0 \\
\lambda_{\rm eh}^0
\end{array}
\right), 
\nonumber\\
\label{lambdaalpha_supp}
\end{eqnarray}
is satisfied in the steady state, where $E$ is the (real) condensate emission energy. 

Here, we prove our claim in the main text: Whenever an exceptional point (EP) $\Omega=0$ of the matrix $\hat A$, where the two eigenvectors $\bm u_\pm$ and eigenvalues $E_\pm$ coalsce, is found, 
there exists a phase boundary in its vicinity that ends at that point.
In the proof below, it is assumed that the matrix $\hat A$ is a smooth function of the input parameters and has maximum of one solution per solution type. 
We emphasize that the existence of EP is crucially due to the non-Hermitian nature of matrix $\hat A$, since a Hermitian matrix always has orthogonal eigenvectors, two of which may never coalesce.

The central quantity for the proof is a complex splitting,
\begin{eqnarray}
\Lambda=\Omega^2=\varphi^2+4\tilde g_0^* g_0
\end{eqnarray}
which is directly related to the difference between the two eigenvalues $E_\pm$ 
(${\rm Re}E_+\ge {\rm Re}E_-$)
given by, 
\begin{eqnarray}
E_{\pm}
=\frac{1}{2}[h_{\rm cav}+h_{\rm eh}\pm \Omega],
\label{Epm}
\end{eqnarray}
with ${\rm Re}\Omega\ge 0$.
As depicted in Fig. \ref{fig_Lambda_supp}(a), we divide $\Lambda$ into regions I-IV in terms of the sign of the imaginary part of $\Lambda$ and whether the system is in the weak- (strong-) coupling regime, i.e.,  $\tilde\delta^2+4{\rm Re}[\tilde g_0^* g_0]< 4\kappa^2(\ge 4\kappa^2)$,  
where $\tilde\delta={\rm Re}\varphi$.
As shown soon below, 
EP satisfies $\tilde\delta^2+4{\rm Re}[\tilde g_0^* g_0]=4\kappa^2$ and at least in the vicinity of ${\rm Im}\Lambda=0$, 
the strong-couping regime lies at ${\rm Re}\Lambda \ge 0$.

We prove the above by showing that the matrix $\hat A$ satisfies the following three properties:
\begin{enumerate}
  \item Only the ``$+(-)$''-solution can arise in region II (III).
  \item Sweeping $\Lambda$ from region III to II across the dotted line in Fig. \ref{fig_Lambda_supp}(a)
  (${\rm Re}\Lambda < 0$ and ${\rm Im}\Lambda=0$) changes the solution type from ``$-$'' to ``$+$'' without discontinuity in the emission energy $E$, resulting in a smooth crossover.
  \item In contrast, when sweeping parameters in a route where $\Lambda$ encircles the EP as III$\rightarrow$IV$\rightarrow$I$\rightarrow$II, there must exist a point where the solution type switches discontinuously, resulting in a phase transition.
\end{enumerate}
From the assumption that $\hat A$ is a smooth function of the input parameters, these properties result in a phase boundary that ends at the EP, proving our claim.

\begin{widetext}

\begin{figure}
\begin{center}
\includegraphics[width=0.55\linewidth,keepaspectratio]{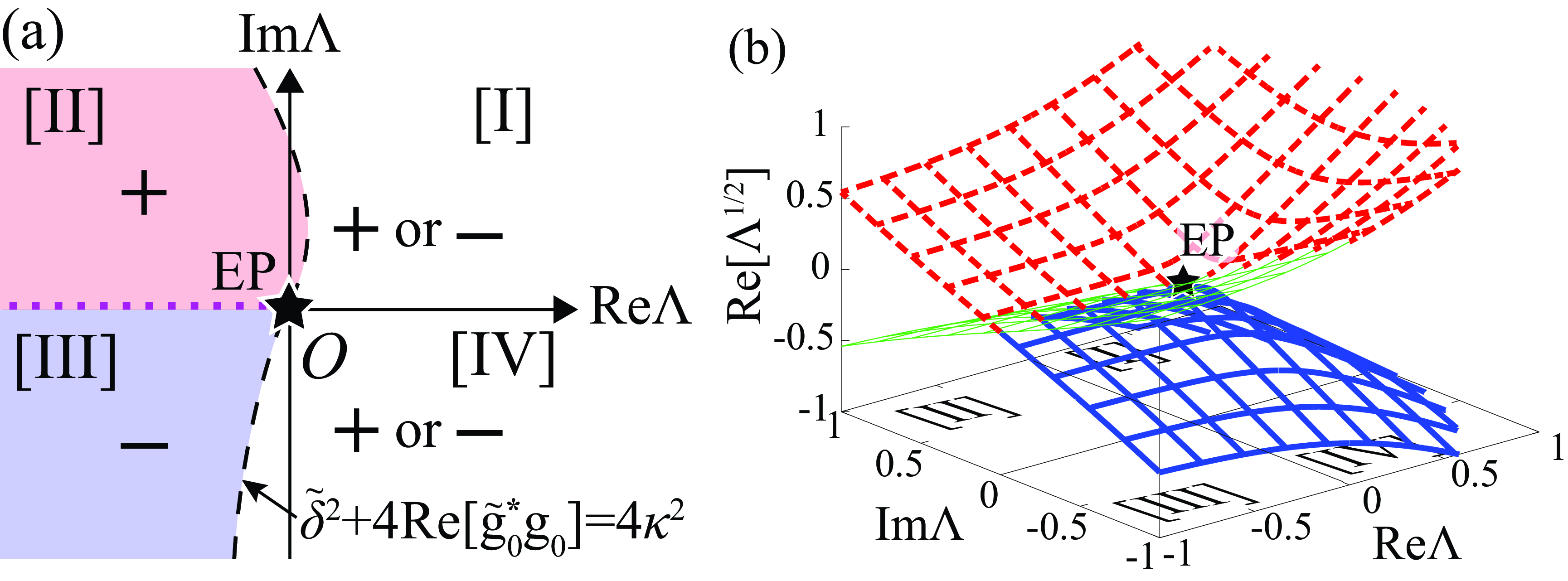}
\end{center}
\caption{(Color online) (a) Definition of regions I-IV. 
In region II (III), only ``$+(-)$''-solution can be realized (property 1).
(b) Plot of the real part of $\sqrt\Lambda$. 
The blue solid and the red dashed line represent different Riemann sheets, where the branch cut lies at ${\rm Re}\Lambda<0$ and ${\rm Im}\Lambda=0$ (the dotted line in panel (a)).
The Riemann surface depicted in thin green lines is the sheet we do not use, due to the restriction from the property 1. 
}
\label{fig_Lambda_supp}
\end{figure}

Let us first prove the property 1. 
The ``$-$'' and ``$+$''-solutions satisfy 
\begin{eqnarray}
0&=&E_- -E
=\frac{1}{2}
\big[
2\xi-i(\kappa-R_{\rm eh})
-\Omega
\big],
\label{E-}
\\
0&=&E_+ -E
=\frac{1}{2}
\big[
2\xi-i(\kappa-R_{\rm eh})
+\Omega
\big],
\label{E+}
\end{eqnarray}
respectively, where
\begin{eqnarray}
\xi&=&\frac{1}{2}\big[{\rm Re}[h_{\rm cav}]+{\rm Re}[h_{\rm eh}]\big]-E,
\\
R_{\rm eh}&=&{\rm Im}h_{\rm eh},
\end{eqnarray}
and
\begin{eqnarray}
\Omega=\sqrt{\Lambda}=
\sqrt{\tilde\delta^2 -(\kappa+R_{\rm eh})^2
+ 4{\rm Re}[\tilde g_0^*g_0] 
-2i\big[\tilde\delta(\kappa+R_{\rm eh})-2{\rm Im}[\tilde g_0^*g_0] \big]}.
\label{Omega}
\end{eqnarray}

Since we have taken ${\rm Re}\Omega\ge 0$ among the two quantities that $\sqrt{\Lambda}$ takes,
from the real part of Eqs. (\ref{E-}) and (\ref{E+}),  
the ``$-(+)$''-solution has $\xi> 0(\le0)$ since we have defined ${\rm Re}\Omega\ge 0$.

Equations (\ref{E-}) and (\ref{E+}) both satisfy,
\begin{eqnarray}
\big[4\xi^2 - (\kappa-R_{\rm eh})^2\big] -4i(\kappa-R_{\rm eh}) \xi
=\Lambda,
\end{eqnarray}
or
\begin{eqnarray}
\xi^2 
&=& 
\frac{1}{4}[(\kappa-R_{\rm eh})^2+{\rm Re}\Lambda]
=\frac{1}{4}
\Big[
-4\kappa R_{\rm eh} +\tilde\delta^2+ 4{\rm Re}[\tilde g_0^*g_0]
\Big], 
\label{ReLambdaxi}
\\
4(\kappa-R_{\rm eh}) \xi
&=& -{\rm Im}\Lambda,
\label{ImLambdaxi}
\end{eqnarray}
\end{widetext}
where we have used, 
\begin{eqnarray}
{\rm Re}\Lambda 
&=&\tilde\delta^2+4{\rm Re}[\tilde g_0^* g_0]-(\kappa+R_{\rm eh})^2,
\label{ReLambda}
\end{eqnarray}
in the second equality of Eq. (\ref{ReLambdaxi}).
Equation (\ref{ImLambdaxi}) gives,
\begin{eqnarray}
{\rm sgn}[\kappa-R_{\rm eh}]{\rm sgn}[\xi] = -{\rm sgn}[{\rm Im}\Lambda],
\end{eqnarray}
telling us that the sign of ${\rm Im}\Lambda$ affects either the magnitude relation of $\kappa$ and $R_{\rm eh}$, or the solution type determined by the sign of $\xi$.

In the weak-coupling regime (regions II and III) $\tilde\delta^2+4{\rm Re}[\tilde g_0^*g_0]<4\kappa^2$, from Eq. (\ref{ReLambdaxi}),
\begin{eqnarray}
R_{\rm eh}&=&
\frac{1}{4\kappa}
\big[
\tilde\delta^2 + 4{\rm Re}[\tilde g_0^*g_0]
-4\xi^2
\big]
\nonumber\\
&\le &
\frac{1}{4\kappa}
\big[
\tilde\delta^2 + 4{\rm Re}[\tilde g_0^*g_0]
\big]
<\kappa,
\label{Reh_weak}
\end{eqnarray}
where we have used $\kappa>0$.
As a result, we get
\begin{eqnarray}
{\rm sgn}[\xi] = -{\rm sgn}[{\rm Im}\Lambda],
\end{eqnarray}
proving that only the ``$-(+)$''-solution given by $\xi>0(\le 0)$ can be realized in region III (II).

We can now show that the EP satisfies, 
\begin{eqnarray}
\tilde\delta^2+4{\rm Re}[\tilde g_0^*g_0]=4\kappa^2,
\label{weak_strong_EP}
\end{eqnarray}
as schematically drawn in Fig. \ref{fig_Lambda_supp}(a).
This follows from the properties that we get $\xi =0$ at ${\rm Re}\Omega=0$ and
$\kappa=R_{\rm eh}$ at ${\rm Im}\Omega=0$
(which may readily be seen from Eqs. (\ref{E-}) and (\ref{E+})), as well as the property that ${\rm Re}\Lambda$ 
vanishes at the EP ($\Omega=0$). 
In addition, we can also show that when
${\rm Im}\Lambda=0$ with ${\rm Re}\Lambda \ge 0$ giving ${\rm Im}\Omega=0$, the state is in the strong-coupling regime, because
it satisfies $\kappa = R_{\rm eh}$, and thus from Eq. (\ref{ReLambda}),
\begin{eqnarray}
0\le {\rm Re}\Lambda = \tilde\delta^2 + 4{\rm Re}[\tilde g_0^*g_0]-4\kappa^2,
\end{eqnarray}
which is our definition of the strong-coupling regime (I and IV).

We next show the properties 2 and 3. 
These properties can be understood from the plot of $\sqrt{\Lambda}$, depicted in Fig. \ref{fig_Lambda_supp}(b).
As seen in the figure, $\sqrt{\Lambda}$ is in general a two-valued quantity, consisting of two Riemann sheets (i.e., the sheets drawn with blue solid lines and red dashed lines).
Noting our definition that ${\rm Re}E_+\ge {\rm Re}E_-$, or ${\rm Re}\Omega\ge 0$,
it can be seen from Eq. (\ref{Epm}) that the Riemann sheet with ${\rm Re}[\sqrt\Lambda]<0(\ge 0)$, depicted with blue solid lines (red dashed lines) in Fig. \ref{fig_Lambda_supp}(b), is used in computing the emission energy $E$ of the ``$-(+)$''-solution.

From the restriction  of the solution types in regions II and III (property 1), we can ignore  the Riemann sheet depicted with thin green lines in Fig. \ref{fig_Lambda_supp}(b).
Since the two Riemann sheets that are used for ``$-$'' and ``$+$''-solutions are connected at the boundary between regions II and III (i.e., the dotted line in Fig. \ref{fig_Lambda_supp}(a)),
the solution types can switch continuously by passing through that boundary, proving the property 2.
Conversely, since that boundary is the only place that connects the two Riemann sheets, it is otherwise associated with exhibiting a discontinuity in physical quantities. 
This proves the property 3, and therefore the theorem.

\section{Hartree-Fock-Bogoliubov approximation}

Here, we show that the dilute equilibrium limit reduces to the conventional polariton condensate picture, 
by showing within the Hartree-Fock-Bogoliubov approximation (HFBA) \cite{Hanai2018_supp,Yamaguchi2012_supp,Yamaguchi2013_supp,Yamaguchi2015_supp} that
the matrix $\hat A$ (Eq. (\ref{lambdaalpha_supp})) is given by [Eq. (6) in the main text],
\begin{eqnarray}
\hat A _{\rm BEC}
=
\left(
\begin{array}{cc}
\hbar\omega_{\rm cav}  & g_{\rm R}  \\
g_{\rm R}^* & \hbar\omega_{\rm X}
\end{array}
\right),
\label{ABEC_supp}
\end{eqnarray}
in this limit  ($\kappa=0,\gamma\rightarrow 0^+,n_{\bm k,\sigma}\ll 1$).
We briefly note that the matrix $\hat A_{\rm BEC}$ is Hermitian in this limit.

The interaction part of the self-energy $\hat\Sigma_{\rm int}$ within HFBA is given by, 
\begin{eqnarray}
\hat{ \Sigma}_{\rm HFB}^{\rm R}(\bm k, \omega;\bm r,t) 
&=&
-\left(
\begin{array}{cc}
0 & \Delta_{\bm k}(\bm r,t) \\
\Delta_{\bm k}^*(\bm r,t)  & 0
\end{array}
\right),
\label{SigHFB_R}
\\
\hat{ \Sigma}_{\rm HFB}^{\rm A}(\bm k, \omega;\bm r,t) 
&=&
-\left(
\begin{array}{cc}
0 & \Delta_{\bm k}(\bm r,t) \\
\Delta_{\bm k}^*(\bm r,t)  & 0
\end{array}
\right),
\label{SigHFB_A}
\\
\hat{ \Sigma}_{\rm HFB}^{\rm K}(\bm k, \omega;\bm r,t) 
&=&
\hat{ \Sigma}_{\rm HFB}^<(\bm k, \omega;\bm r,t) 
=0.
\label{SigHFB_K}
\end{eqnarray}
We refer to Refs. \cite{Hanai2018_supp,Yamaguchi2012_supp,Yamaguchi2013_supp,Yamaguchi2015_supp} for the derivation.
This yields,
\begin{eqnarray}
I_{\bm k,{\rm e}}(\bm r,t)
&=&
\frac{2\gamma_{\rm e}}{\hbar}
n_{\bm k,{\rm e}}^{\rm env}(\bm r,t)
-2{\rm Im}[\Delta_{\bm k}p_{\bm k}^*],
\\
I_{\bm k,{\rm h}}(\bm r,t)
&=&
\frac{2\gamma_{\rm h}}{\hbar}
n_{\bm k,{\bm h}}^{\rm env}(\bm r,t)
-2{\rm Im}[\Delta_{\bm k}p_{\bm k}^*],
\\
I_{\bm k}^{\rm pol}(\bm r,t)
&=&
\frac{1}{\hbar}
\big[2\gamma p_{\bm k}^{\rm env}(\bm r,t)
+i\Delta_{\bm k}(\bm r,t)N_{\bm k}(\bm r,t)\big],
\nonumber\\
\end{eqnarray}
where $N_{\bm k}(\bm r,t)=1-n_{\bm k,{\rm e}}(\bm r,t)-n_{\bm k,{\rm h}}(\bm r,t)$ is the population inversion.

In the dilute equilibrium limit ($\kappa=0,\gamma\rightarrow 0^+,N_{\bm k}\simeq 1$) in the steady state, 
the term $I_{\bm k}^{\rm pol}(\bm r,t)$ simplifies to,
\begin{eqnarray}
&&I_{\bm k}^{\rm pol}(\bm r,t)
=\frac{i}{\hbar}\Delta_{\bm k}(\bm r,t)
\nonumber\\
&&=\frac{i}{\hbar}
\Big[\sum_{\bm k'}V_{\bm k-\bm k'}p_{\bm k'}^0
-g\lambda_{\rm cav}^0\Big]
e^{-iEt/\hbar},
\end{eqnarray}
which gives $L^{\rm eq,dil}_{\bm k,\bm k'}=\delta_{\bm k,\bm k'}$.
In this case, from Eq. (\ref{pk_Boltzmann}), 
\begin{eqnarray}
\Big(
\frac{\hbar^2\bm k^2}{m_{\rm eh}}+(E_{\rm g}-E)
\Big)
\lambda_{\rm eh}^0\phi_{\bm k}
= \sum_{\bm k'}V_{\bm k-\bm k'}\phi_{\bm k'}
\lambda_{\rm eh}^0
-g\lambda_{\rm cav}^0.
\nonumber\\
\label{pk_Boltzmann_eq}
\end{eqnarray}
For simplicity, let us assume that 
\begin{eqnarray}
\sum_{\bm k'}V_{\bm k-\bm k'}\lambda_{\rm eh}^0\phi_{\bm k'}
\gg g\lambda_{\rm cav}^0.
\label{assumption}
\end{eqnarray}
This assumption implies $g_{\rm R}\ll E_{\rm X}^{\rm bind}$
(where $g_{\rm R}$ and $E_{\rm X}^{\rm bind}$ are the Rabi splitting and the exciton binding energy, respectively), as shown soon later.
In this situation, Eq. (\ref{pk_Boltzmann_eq}) 
reduces to the Schr\"odinger equation of an exciton,
\begin{eqnarray}
\frac{\hbar^2\bm k^2}{m_{\rm eh}}
\phi^{\rm X}_{\bm k}
-\sum_{\bm p}V_{\bm k-\bm k'}\phi^{\rm X}_{\bm k'}
=-E^{\rm bind}_{\rm X}\phi^{\rm X}_{\bm k},
\label{Schrodinger}
\end{eqnarray}
or
\begin{eqnarray}
&&\int d\bm r' \Big[
\delta(\bm r-\bm r')\frac{-\hbar^2\nabla'^2}{m_{\rm eh}}
-V(\bm r-\bm r')
\Big]\phi_{\rm X}(\bm r')
\nonumber\\
&&\ \ \ \ \ \ \ \ \ \ \ \ \ \ \ \ 
=-E_{\rm X}^{\rm bind}\phi_{\rm X}(\bm r),
\end{eqnarray}
where 
$\phi_{\bm k}=\phi^{\rm X}_{\bm k}$ is the exciton wave function.
Here, note that 
\begin{eqnarray}
 |E_{\rm g}-E| - E_{\rm X}^{\rm bind} \ll E_{\rm X}^{\rm bind},
 \label{EgE}
\end{eqnarray}
needs to be satisfied
for Eqs. (\ref{pk_Boltzmann_eq}) and (\ref{Schrodinger})  to be compatible under the assumption (\ref{assumption}).
As a result, the off-diagonal components in the matrix $\hat A$
reduces to the Rabi splitting $g_{\rm R}$,
\begin{eqnarray}
g_0\simeq \tilde g_0
\simeq g\phi_{\rm X}(\bm r=0)
= g_{\rm R},
\end{eqnarray}
and
\begin{eqnarray}
&&h_{\rm eh}^{\rm eq,dil}
\simeq
\sum_{\bm k}
\Big[
\Big(
\frac{\hbar^2\bm k^2}{m_{\rm eh}}+E_{\rm g}
\Big)
|\phi^{\rm X}_{\bm k}|^2
-\sum_{\bm k'}V_{\bm k-\bm k'}\phi^{\rm X}_{\bm k'}
\phi_{\bm k}^{\rm X*}
\Big]
\nonumber\\
&&=
\sum_{\bm k}
\Big[
\frac{\hbar^2\bm k^2}{m_{\rm eh}}
\phi^{\rm X}_{\bm k}
-\sum_{\bm k'}V_{\bm k-\bm k'}\phi^{\rm X}_{\bm k'}
\Big]\phi_{\bm k}^{\rm X*}
+E_{\rm g}
\nonumber\\
&&=-E^{\rm bind}_{\rm X}\sum_{\bm k}|\phi^{\rm X}_{\bm k}|^2+E_{\rm g}
=E_{\rm g}-E^{\rm bind}_{\rm X}
=\hbar\omega_{\rm X},
\end{eqnarray}
which yields the desired Eq. (6). 
Since $|E_{\rm g}-E|- E_{\rm X}^{\rm bind}=|E_{\rm g}-E_{\rm LP/UP}|- E_{\rm X}^{\rm bind}\sim g_{\rm R}$ unless the system is not in an extreme red or blue detuning (where $E_{\rm LP(UP)}$ is the lower (upper) polariton energy), from Eq. (\ref{EgE}), our assumption (\ref{assumption}) is satisfied at $g_{\rm R}\ll E_{\rm X}^{\rm bind}$.


\section{Driven-dissipative Gross-Pitaevskii Equation (7)}

Let us turn to the polariton laser regime, where the nonequilibrium condensate is dilute enough such that the polariton picture still holds, 
and show that the matrix $\hat A$ in this regime is given by the driven-dissipative Gross-Pitaevskii (ddGP) equation \cite{Wouters2007_supp}
(Eq. (7) in the main text),
\begin{eqnarray}
\hat A_{\rm GP}=
\left(
\begin{array}{cc}
\hbar\omega_{\rm cav}-i\kappa & g_{\rm R}  \\
g_{\rm R}^* & \hbar\omega_{\rm X}+U_{\rm X}|\lambda_{\rm eh}^0|^2 + i R_{\rm X}
\end{array}
\right).
\label{GP_supp}
\end{eqnarray} 
In this regime, $\Delta L_{\bm k,\bm k'}\equiv L_{\bm k,\bm k'}-L^{\rm eq,dil}_{\bm k,\bm k'}$
that characterizes the derivation from the dilute equilibrium limit
is small ($|\Delta L_{\bm k,\bm k'}|\ll 1$).
In this case, under the assumption (\ref{assumption}), we may approximate $\phi_{\bm k}$ as 
$\phi_{\bm k}\simeq \phi_{\bm k}^{\rm X}$ giving $g_0\simeq g_{\rm R}$
and
\begin{eqnarray}
&&E\lambda_{\rm eh}^0 =\tilde g_0^* \lambda_{\rm cav}^0 
+ h_{\rm eh}\lambda_{\rm eh}^0 
\nonumber\\
&&\simeq 
g_{\rm R}\lambda_{\rm cav}^0
+\big(\hbar\omega_{\rm X}
-\sum_{\bm k,\bm k',\bm p}
 V_{\bm k-\bm p}\phi_{\bm k}^*\phi_{\bm p}
 \Delta L_{\bm k,\bm k'}
 \big) \lambda_{\rm eh}^0
 \nonumber\\
&&\equiv
g_{\rm R}\lambda_{\rm cav}^0
+\big(\hbar\omega_{\rm X}+\Delta U+
iR_{\rm X}
\big)\lambda_{\rm eh}^0.
\label{second_polariton}
\end{eqnarray}
Here, $\Delta U$ and $R_{\rm X}$ physically describe the blue shift of the exciton spectrum and the exciton gain, respectively.
Expanding $\Delta U$ in terms of $|\lambda_{\rm eh}^0|^2$ and neglecting the blue shift from the non-coherent part,  
\begin{eqnarray}
\Delta U\simeq U_{\rm X}|\lambda_{\rm eh}^0|^2,
\end{eqnarray}
we obtain the desired matrix $\hat A_{\rm GP}$ of the driven-dissipative Gross-Pitaevskii equation (7).

\section{Vertical-cavity surface-emitting laser (VCSEL) regime}

We show here some properties of the solution types realized in the high density region, where the system operates as a vertical-cavity surface-emitting laser (VCSEL).
The VCSEL regime schematically depicted in Fig. 3(b) in the main text is based on the properties analyzed here.
In this regime, as mentioned in the main text, the dynamics of this system is given by the semiconductor Maxwell-Bloch equations and the matrix $\hat A$ is given by Eq. (8),
\begin{eqnarray}
\hat A_{\rm VL}=
\left(
\begin{array}{cc}
\hbar\omega_{\rm cav}-i\kappa & g_0  \\
\tilde g_0^{\rm VL*}
& \hbar\omega_{\rm eh}^{\rm VL}-2i\gamma
\end{array}
\right),
\end{eqnarray}
where $\hbar\omega_{\rm eh}^{\rm VL}
=\sum_{\bm k}[(\varepsilon_{\bm k,{\rm e}}+\varepsilon_{\bm k,{\rm h}})|\phi_{\bm k}|^2 
- \sum_{\bm p}V_{\bm k-\bm p}\phi_{\bm k}^*\phi_{\bm p}N_{\bm k}]$ and $\tilde g_0^{\rm VL*}=g\sum_{\bm k}\phi_{\bm k}^*N_{\bm k}$.
In this case, the eigenvalues are given by,
\begin{eqnarray}
E_\pm^{\rm VL} = \frac{1}{2}[(\hbar\omega_{\rm cav}+\hbar\omega_{\rm eh}^{\rm VL})-i(\kappa+2\gamma)\pm\Omega_{\rm VL}],
\end{eqnarray}
with
\begin{eqnarray}
\Omega_{\rm VL}\simeq\sqrt{\delta_{\rm VL}^2-4|g_0|^2-2i\delta_{\rm VL}(\kappa-2\gamma)}.
\end{eqnarray}
Here, we have assumed that the system is in an extremely strong pumping regime such that a large population inversion $N_{\bm k}\simeq -1$ exists at a predominant momentum window, 
which makes 
$\hbar\omega_{\rm eh}^{\rm VL}\simeq\sum_{\bm k}[(\varepsilon_{\bm k,{\rm e}}+\varepsilon_{\bm k,{\rm h}})|\phi_{\bm k}|^2 
+ \sum_{\bm p}V_{\bm k-\bm p}\phi_{\bm k}^*\phi_{\bm p}$ a real number and $\tilde g_0^{\rm VL*}=-g_0^*$.

In this regime, since $E$ needs to be real, 
\begin{eqnarray}
\kappa+2\gamma \simeq  {\rm Im}\Omega_{\rm VL} (-{\rm Im}\Omega_{\rm VL}),
\label{ErealVL_supp}
\end{eqnarray}
for the ``$+(-)$''-solution. 
From this relation, we can conclude that the ``$+(-)$''-solution is realized when ${\rm Im}\Lambda_{\rm VL}>0(<0)$  (where $\Lambda_{\rm VL}=\Omega_{\rm VL}^2$), which implies that the solution type depends strongly on the details of the experimental setup.

In addition, again from Eq. (\ref{ErealVL_supp}), ${\rm Re}\Lambda_{\rm VL}<0$ is satisfied when ${\rm Im}\Lambda_{\rm VL}=0$,
stating that the VCSEL is in the weak-coupling regime when $\delta_{\rm VL}\simeq0$ or $\kappa\simeq 2\gamma$.
This can be shown as follows: Let us assume that ${\rm Re}\Lambda_{\rm VL}\ge 0$  when ${\rm Im}\Lambda_{\rm VL}=0$. 
Then, $\Omega_{\rm VL}$ is real and non-negative, i.e., ${\rm Im}\Omega_{\rm VL}=0$ because ${\rm Re}\Omega_{\rm VL}\ge 0$ according to our definition. 
This, however, contradicts with Eq. (\ref{ErealVL_supp}).
On the other hand, ${\rm Re}\Lambda_{\rm VL}< 0$ gives a pure imaginary $\Omega_{\rm VL}$ as it is supposed to, proving the above claim.

\begin{widetext}

\begin{figure}
\begin{center}
\includegraphics[width=0.8\linewidth,keepaspectratio]{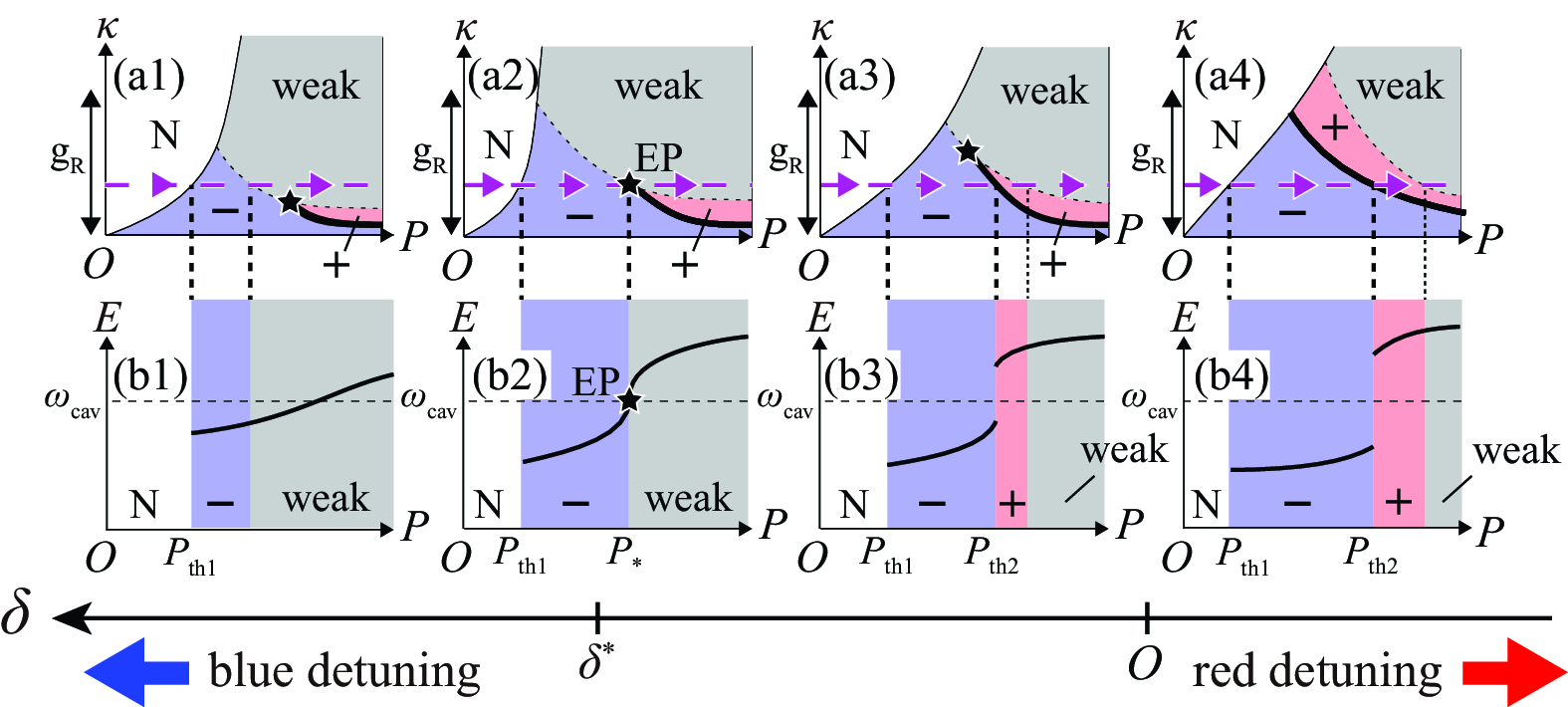}
\end{center}
\caption{(Color online) Proposed phase diagram of a driven-dissipative electron-hole-photon gas (a1)-(a4) 
and the expected (schematic) pump power $P$ dependence with fixed photon decay rate $\kappa(<g_{\rm R})$ (i.e., the dashed arrows in (a1)-(a4)) of 
the condensate emission $E$ (b1)-(b4) from our theory.
Here, each column of panels represent our prediction at different detuning $\delta$, where we put the bluer (reder) detuning on the left (right).
Panels (a2) and (b2) are at a critical detuning $\delta=\delta^*$ where the state passes through the EP at a fixed $\kappa$ (dashed arrow). 
In panels (a1)-(a4), the blue (red) shaded region labeled by ``$-(+)$'' is the condensed ``$-(+)$''-solution phase in the strong-coupling regime. 
The gray shaded area represents the condensed phase in the weak-coupling regime, where ``$-$'' and ``$+$''-solution phase crosses over to each other (for this reason, we omit the ``$-$'' and ``$+$'' labeling in this regime). 
The solid line represents the first-order-like phase boundary between the ``$-$'' and ``$+$''-solution phase, and the star represents the exceptional point.  
``N'' is the normal phase and the thin line represents the phase boundary between the normal and the condensed phase. 
In panels (b1)-(b4), $P_{\rm th1}$ and $P_{\rm th2}$ are the critical pump power $P$ at the first and the second threshold, respectively, and $P=P_*$ is the pump power at the EP.  
}
\label{fig_experiment_supp}
\end{figure}
\end{widetext}

\section{Implications for experiments}

We address here, in detail, the experimental consequences of our scenario, which we find in qualitative agreement with the existing data. 
We also propose a possible experiment that may be employed to test our theory more directly. 

\subsection{Comparison to Experiments}

Figure \ref{fig_experiment_supp} summarizes our expectation from our theory. 
As we argue in detail below, we predict a single-threshold-behavior to a photon laser at a large blue detuning $\delta\gg 0$ (Fig. \ref{fig_experiment_supp} (b1)),
as a function of the pump power $P$, 
up to the detuning $\delta=\delta^*(>0)$ where the system passes through the EP (Fig. \ref{fig_experiment_supp} (b2)). Beyond this point, i.e., at more red detuning $\delta<\delta^*$, we expect a two-threshold-behavior (Figs. \ref{fig_experiment_supp} (b3), (b4)). Here, the first threshold $P=P_{\rm th1}$ is attributed to the normal-to-condensate transition, while the second threshold $P=P_{\rm th2}$ is attributed to the first-order-like phase transition from the ``$-$'' to ``$+$''-solution phase transition associated with a discontinuity in the condensate emission energy $E$. 
These predictions are consistent with experiments, where most experiments that reports the existence of the second threshold are in red detuning or on resonance 
\cite{Bajoni2008_supp,Balili2009_supp,Nelsen2009_supp,Tempel2012a_supp,Tempel2012b_supp,Tsotsis2012_supp,Horikiri2013_supp,Fischer2014_supp,Kim2016_supp,Brodbeck2016_supp,Schneider2013_supp},
and a single-threshold-behavior to a photon laser is reported at a large blue detuning \cite{Deng2003_supp}.

The second threshold has traditionally been interpreted as a signal of a strong-to-weak-coupling transition, i.e., a polariton laser to a photon laser transition. 
However, our theory provides a possible new interpretation to this phenomenon, i.e., a lower-to-upper-branch condensate transition 
(strong-to-\textit{strong}-coupling transition).
This scenario is supported by several experiments showing a small but relevant \textit{blue} shift of the condensate emission energy $E$ from the cavity mode $\hbar\omega_{\rm cav}$ (i.e., $E>\hbar\omega_{\rm cav}$) \cite{Tempel2012a_supp,Tempel2012b_supp,Fischer2014_supp,Kim2016_supp}, 
just above the second threshold $P\gesim P_{\rm th2}$.
These observations are consistent with our picture of upper-branch condensation that may arise above the second threshold $P>P_{\rm th2}$ (Figs. \ref{fig_experiment_supp}(b3), (b4)) which can have an emission energy above the cavity mode energy, $E=E_+>\hbar\omega_{\rm cav}$, due to the Rabi splitting (which, in the high density region, would be substantially reduced from the bare Rabi splitting $g_{\rm R}$ due to the phase filling effect).
In contrast, they do not agree with a conventional photon laser picture, where a \textit{red} shift from the cavity mode $\hbar\omega_{\rm cav}$ (i.e., $E<\hbar\omega_{\rm cav}$) is usually obtained due to the mode-pulling effect \cite{Haug_supp}. 
These observations, combined with our theory, strongly imply that the observation of the second threshold alone cannot be identified as a signal of a polariton to photon laser transition.

\begin{figure}
\begin{center}
\includegraphics[width=0.45\linewidth,keepaspectratio]{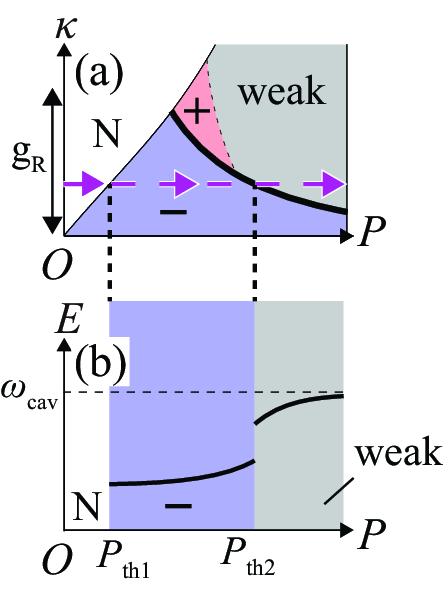}
\end{center}
\caption{(Color online) Possible phase diagram for red detuning $\delta<0$ (a) and the condensate emission energy $E$ (b). In this case, a strong-to-weak-coupling transition takes place at the second threshold $P=P_{\rm th2}$. 
The meaning of ``N'', ``$-$'', ``$+$'', and ``weak'' are the same as in Fig. \ref{fig_experiment_supp}.}
\label{fig_experiment_strong_weak_supp}
\end{figure}

We briefly note that our theory does not exclude the conventional strong-to-weak-coupling transition scenario. 
The ``$-$''-solution in the strong-coupling regime may exhibit a phase transition to the ``$+$''-solution in the \textit{weak}-coupling regime,  
as depicted in Fig. \ref{fig_experiment_strong_weak_supp}. 
Indeed, many experiments show condensate emission energy $E$ below $\hbar\omega_{\rm cav}$ at $P\approx P_{\rm th2}$ 
\cite{Bajoni2008_supp,Balili2009_supp,Nelsen2009_supp,Horikiri2013_supp,Brodbeck2016_supp,Schneider2013_supp}
consistent with the conventional photon laser picture.
Furthermore, our theory does not exclude the possibility of a weak-to-weak-coupling transition scenario.  
Unfortunately, our general framework cannot determine which scenario actually occurs for a given experimental setup 
and requires a concrete computation based on approximations \cite{Hanai2018_supp,Haug_supp,Kwong2009_supp,Asano2014_supp}, which remains as our future work.

Below, we provide a more detailed discussion of these claims.
Let us start by considering the situation where the EP is found in the dilute (polariton laser) regime, 
which turns out to be the case at detunings close to resonance. 
As shown earlier in this Supplemental Material, 
the equation of motion in this regime is determined by the ddGP equation (\ref{GP_supp}) (Eq. (7) in the main text). 
In this case, we find the EP at
\begin{eqnarray}
\tilde\delta &=& \delta - U_{\rm X}|\lambda_{\rm eh}^0|^2 = 0,
\label{deltaEP_GP_supp}
\\
\kappa &=& R_{\rm X}=g_{\rm R}.
\label{kappaEP_GP_supp}
\end{eqnarray}
In the red detuning $\delta<0$, from Eq. (\ref{deltaEP_GP_supp}), no EP exists. 
As discussed in the main text, the effective detuning $\tilde\delta$ decreases as the pump power $P$ increases, so that the complex splitting, 
\begin{eqnarray}
\Lambda\simeq\Lambda_{\rm GP}
=\tilde\delta^2+4|g_{\rm R}|^2-(\kappa+R_{\rm X})^2
-2i\tilde\delta(\kappa+R_{\rm X}),
\nonumber\\
\end{eqnarray}
varies counter-clockwise around the EP in $\Lambda$-space as a function of $P$, giving rise to a phase transition from the ``$-$'' to the ``$+$''-solution phase.
(See Fig. \ref{fig_experiment_supp}(a4).) 
On resonance, $\delta=0$, the EP is found at the zero condensate fraction limit $|\lambda_{\rm eh}^0|^2, |\lambda_{\rm cav}^0|^2\rightarrow 0$ with $\kappa=g_{\rm R}$, meaning that the EP is found on the phase boundary between the condensed and the normal phase (as depicted in Fig. 1(b) in the main text). 
At blue detuning $\delta>0$, the EP is found at a finite condensate fraction $|\lambda_{\rm eh}^0|^2 = \delta / U_{\rm X}>0$, 
giving rise to an endpoint to the first-order-like phase boundary (Fig. \ref{fig_experiment_supp}(a1)-(a3)).

As derived in Eqs. (\ref{deltaEP_GP_supp}) and (\ref{kappaEP_GP_supp}), according to the ddGP equation (\ref{GP_supp}), the  EP always lies at $\kappa=g_{\rm R}$ at arbitrarily blue detuning. 
However, the ddGP equation (\ref{GP_supp}) is only valid in the dilute limit, which can only be true at detuning not very far away from resonance 
with low enough bulk temperature.
As one moves the detuning deeper into the blue detuning regime $\delta\gg 0$, the EP shifts to higher density, until the the phase filling effect starts to come into play and the ddGP equation (\ref{GP_supp}) becomes invalid. 
In this case, we argue below that the phase filling effect shifts the EP to a smaller photon decay rate $\kappa(<g_{\rm R})$.
(See Fig. \ref{fig_experiment_supp}(a1)-(a3).)

The phase filling effect can be taken into account by considering a slightly modified version of the ddGP equation (\ref{GP_supp}),
 \begin{eqnarray}
\hat A_{\rm GP}^{\rm pfe}=
\left(
\begin{array}{cc}
\hbar\omega_{\rm cav}-i\kappa & g_{\rm R}  \\
\tilde g_0^* & \hbar\omega_{\rm X}+U_{\rm X}|\lambda_{\rm eh}^0|^2 + i R_{\rm X}
\end{array}
\right),
\label{GP_pfe_supp}
\end{eqnarray} 
where one of the off-diagonal components is replaced by an effective Rabi splitting suppressed by the phase filling effect, 
\begin{eqnarray}
\tilde g_0^*= g\sum_{\bm k}\phi^{\rm X*}_{\bm k} N_{\bm k}.
\end{eqnarray}
Here, the finite electron-hole density that results in 
$N_{\bm k}=1-n_{\bm k,{\rm e}}-n_{\bm k,{\rm h}} < 1$
suppresses the effective Rabi splitting $|\tilde g_0^*| < g_{\rm R}$. 
This modifies the EP to lie at 
\begin{eqnarray}
\kappa \simeq R_{\rm X}
\simeq \sqrt{[{\rm Re}\tilde g^*_0]g_{\rm R}}
<g_{\rm R},
\end{eqnarray}
and $\tilde\delta=\delta - U_{\rm X}|\lambda_{\rm eh}^0|^2 = 0$, 
where we have neglected the imaginary part of $\tilde g_0^*$ for simplicity.
From this argument, we predict the EP to be found at smaller $\kappa (<g_{\rm R})$ at bluer detunings, as depicted schematically in Figs. \ref{fig_experiment_supp}(a1) and (a2).

This leads us to our prediction above: a single-threshold-behavior at $\delta>\delta^*$ and a two-threshold-behavior at $\delta<\delta^*$. 
This can be seen by noting that, typically, experiments study the pump power $P$ dependence in a cavity with a fixed photon decay rate 
(typically $\kappa=0.01{\rm meV}-1{\rm meV}$) which is smaller than the Rabi splitting ($g_{\rm R}\sim 5{\rm meV}-10{\rm meV}$ in GaAs). Thus, the parameter changes along the dashed line in Figs. \ref{fig_experiment_supp}(a1)-(a4). 

As seen in the figure, at all detunings, the normal-to-condensate transition takes place at $P=P_{\rm th1}$, where the dashed line intersects with the thin line (the first threshold). 
In addition, we find another intersection between the dashed and the solid line at $\delta<\delta^*$, which is nothing but the second threshold, $P=P_{\rm th2}$ (Figs. \ref{fig_experiment_supp}(a3), (a4)).

Our result suggests a possibility of interpreting the second threshold $P_{\rm th2}$ as a signal of a lower-to-upper branch condensate transition. 
The fact the ddGP equation (\ref{GP_supp}) gives rise to the second threhold despite the property that polariton picture still holds (as demonstrated in Fig. 4 in the main text) supports this scenario.  
On the other hand, our scenario may also lead to the (traditional) strong-to-weak-coupling transition at the second threshold as well. 
In fact, we believe that some of the observed second thresholds \cite{Tempel2012a_supp,Tempel2012b_supp,Fischer2014_supp,Kim2016_supp}
can be interpreted as the former type of transition, while others
\cite{Bajoni2008_supp,Balili2009_supp,Nelsen2009_supp,Horikiri2013_supp,Brodbeck2016_supp,Schneider2013_supp}
are interpreted as the latter, as we discuss below. 

At high pump power $P$ or large photon decay rate $\kappa$,
 the weak-coupling regime, which we have defined as the region that satisfies
\begin{eqnarray}
\tilde \delta^2 + 4{\rm Re}[\tilde g_0^* g_0] < 4\kappa^2,
\end{eqnarray}
is expected to arise, again due to the strong phase filling effect. 
The VCSEL typically lies in this regime.
As proved earlier, the EP lies on the boundary between the weak- and strong-coupling regimes, since the equality $\tilde \delta^2 + 4{\rm Re}[\tilde g_0^* g_0] = 4\kappa^2$ holds at the EP.
Keeping in mind that the complex splitting $\Lambda$ may discontinuously change in between the first-order phase transition,  
we may consider several possible scenarios on where the weak-coupling regime appears, which can lead to different properties at the second threshold. 

One candidate is depicted in Figs. \ref{fig_experiment_supp}(a3), (b3), (a4), and (b4). 
Here, the gray shaded area represents the weak-coupling regime. 
(We have omitted the ``$-$'' and ``$+$'' labeling of the solution type in this regime, since the two types cross over to one another smoothly without discontinuity and thus the labeling is not very important.)
In this phase diagram, a first-order-like transition occurs at $P=P_{\rm th2}$ \textit{within} the strong-coupling regime, i.e., the lower-to-upper branch condensate transition. 
This scenario is consistent with the experiments that report a blue shift of the condensate emission energy $E$ from the cavity mode $\hbar\omega_{\rm cav}$ \cite{Tempel2012a_supp,Tempel2012b_supp,Fischer2014_supp,Kim2016_supp}, as mentioned above. 

Another possibility is depicted in Fig. \ref{fig_experiment_strong_weak_supp}, where the ``$-$''-solution phase in the strong-coupling regime exhibist a phase transition to the weak-coupling regime (``$+$'' solution).
This follows the traditional strong-to-weak-coupling transition interpretation. 

In addition to the above, in principle, a transition from the weak-to-weak-coupling transition is also possible.
The type of transition that actually occurs depends on microscopic details, which its determination needs further analysis \cite{Hanai2018_supp,Haug_supp,Kwong2009_supp,Asano2014_supp}. This remains as our future work.

\subsection{Experimental Proposal}

\begin{figure}
\begin{center}
\includegraphics[width=0.5\linewidth,keepaspectratio]{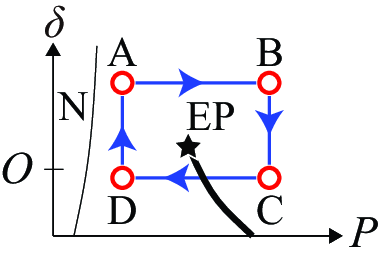}
\end{center}
\caption{(Color online) Proposed experiment to test our theory. Here, by tuning the detuning $\delta$ and the pump power $P$ with a fixed photon decay rate $\kappa(<g_{\rm R})$ in the route A$\rightarrow$B$\rightarrow$C$\rightarrow$D, the state encircles the EP. 
According to our theory, a first-order-like phase transition should take place an odd number of times during the sweep.}
\label{fig_experiment_detuning_supp}
\end{figure}
The main result of our theory is that the phase boundary that gives rise to the second threshold has an endpoint in the blue detuning $\delta>0$. 
Here, we propose that our claim can be tested by tracking the emission energy $E$ and observing that a jump associated with the phase transition occurs an odd number of times when encircling the EP.

To be concrete, we consider a fixed photon decay rate $\kappa(<g_{\rm R})$ as in most experiments.
At a large blue detuning $\delta>\delta^*$, as we have claimed above, we expect a single-threshold-behavior as a function of the pump power $P$ (Fig. \ref{fig_experiment_supp}(a1)), 
which corresponds to sweeping parameters from the point A to B in Fig. \ref{fig_experiment_detuning_supp}. 
On the other hand, a two-threshold-behavior is found at $\delta<\delta^*$ (Fig. \ref{fig_experiment_supp}(a3),(a4)),
corresponding to the sweep from point D to C in Fig. \ref{fig_experiment_detuning_supp}.
Since, according to our theory, there exists an EP that gives an endpoint to the phase boundary in between the two detunings, by tuning both the detuning $\delta$ and the pump power $P$ to move along the arrow in Fig. \ref{fig_experiment_detuning_supp} in the parameter space to encircle the EP, an odd number of first-order-like phase transitions should be observed. 
This should work as an experimental test to our theory. 

Since the proposed scheme only requires \textit{encirclement} the EP, rather than approaching it,   
it should not require fine-tuning of parameters. 
Given that both single- \cite{Deng2003_supp} and two-threshold-behavior \cite{Bajoni2008_supp,Balili2009_supp,Nelsen2009_supp,Tempel2012a_supp,Tempel2012b_supp,Tsotsis2012_supp,Horikiri2013_supp,Fischer2014_supp,Kim2016_supp,Brodbeck2016_supp,Schneider2013_supp},
have been observed, we believe that encircling the EP is possible within current experimental techniques, for instance using wedge cavities in which the detuning varies spatially. 

We finally note that, while the EP appears typically in pairs in conventional non-Hermitian dynamics (See e.g., Ref. \cite{Heiss2004_supp}.), only a single EP is present in our phase diagram, which is crucial for our proposed test. 
This is due to the physical restriction on the matrix $\hat A$ that the decay rate of photons $\kappa$ is always positive. 
If we expand our phase diagram to negative $\kappa$, the partner EP would be found in the unphysical region $\kappa<0$.


\begin{thebibliography}{99}
\bibitem{NOTE_solve} Precisely speaking, our assumption is that the Dyson's equation (where its explicit form is given in the SM \cite{Supp}) is solved to obtain all the terms in Eqs. (\ref{pk})-(\ref{Langevin}).
\bibitem{NOTE_weakstrong} It is shown in the SM \cite{Supp} that the strong-coupling regime $\tilde \delta^2+4{\rm Re}[\tilde g_0^* g_0]\ge 4\kappa^2$ is realized at ${\rm Re}\Lambda \ge 0$, at least in the vicinity of ${\rm Im}\Lambda=0$.
\bibitem{NOTE_theorem} We have implicitly assumed that $\hat A$ is a smooth function of the input parameters and has maximum of one solution per solution type.
\bibitem{NOTE_BEC} In deriving Eq. (\ref{ABEC}), we have assumed $g_{\rm R}\ll E_{\rm X}^{\rm bind}$, for simplicity.
\bibitem{NOTE_GP} The nonlinearity and the gain is present only in the diagonal exciton component under the condition $g_{\rm R}\ll E_{\rm X}^{\rm bind}$ \cite{Supp}.
\bibitem{NOTE_VCSEL} Here, we have assumed that the VCSEL is in an extremely strong pumping regime where a large population inversion $N_{\bm k}\simeq -1$ exists at predominant momentum window, which makes 
$\hbar\omega_{\rm eh}^{\rm VL}
\simeq\sum_{\bm k}[(\varepsilon_{\bm k,{\rm e}}+\varepsilon_{\bm k,{\rm h}})|\phi_{\bm k}|^2 
+ \sum_{\bm p}V_{\bm k-\bm p}\phi_{\bm k}^*\phi_{\bm p}$ 
a real number and $\tilde g_0^{\rm VL*}\simeq-g_0^*$.
\end{thebibliography}
\end{document}